\documentclass{aa}
\usepackage{longtable}
\usepackage{graphicx}
\usepackage{txfonts}
\usepackage{sidecap}

\def\targ{IRAS~21078$+$5211}

\def\nh3{NH$_{3}$}
\def\kms{km~s$^{-1}$}

\def\Vlsr{$V_{\rm LSR}$}
\def\Jyb{Jy~beam$^{-1}$}

\def\G24{G24.78$+$0.08}

\newcommand{\ms}{$M_{\odot}$}
\newcommand{\ls}{$L_{\odot}$}

\newcommand{\pas}{$\rlap{.}^{\prime\prime}$}
\newcommand{\pss}{$\rlap{.}^{\rm s}$}

\newcommand{\degree}{$^{\circ}$}

\usepackage{placeins}

\begin{document}

\title{Protostellar Outflows at the EarliesT Stages (POETS)
\\ V. The launching mechanism of protostellar winds via water masers}
   
  \titlerunning{The launching mechanism of protostellar winds via water masers }

   \author{L. Moscadelli \inst{1}
          \and
           A. Oliva \inst{2,3}
          \and
          A. Sanna \inst{4}
          \and
          G. Surcis \inst{4}
          \and
          O. Bayandina \inst{1}  
          }

   \institute{INAF-Osservatorio Astrofisico di Arcetri, Largo E. Fermi 5, I-50125, Firenze, Italy \\
              \email{luca.moscadelli@inaf.it}
         \and
         Département d'Astronomie, Université de Genève, Chemin Pegasi 51, CH-1290 Versoix, Switzerland
        \and
        Space Research Center (CINESPA), School of Physics, University of Costa Rica, 11501 San Jos\'e, Costa Rica
         \and
         INAF - Osservatorio Astronomico di Cagliari, Via della Scienza 5, 09047 Selargius (CA), Italy
             }

  \abstract
   {Understanding the launching mechanism of winds and jets remains one of the fundamental challenges in astrophysics. The Protostellar Outflows at the EarliesT Stages (POETS) survey has recently mapped the 3D velocity field of the protostellar winds in a sample (37) of luminous young stellar objects (YSOs) at scales of 10--100~au via very long baseline interferometry (VLBI) observations of the 22~GHz water masers. In most of the targets, the distribution of the 3D maser velocities can be explained in terms of a magnetohydrodynamic (MHD) disk wind (DW).}
   {Our goal is to assess the launching mechanism of the protostellar wind in the YSO \targ, the most promising MHD DW candidate from the POETS survey.}
   {We have performed multi-epoch Very Long Baseline Array (VLBA) observations of the 22~GHz water masers in \targ\ to determine the 3D velocities of the gas flowing along several wind streamlines previously identified at a linear resolution of $\sim$1~au.}
   {Near the YSO at small separations along ($xl \le 150$~au) and across ($R \le 40$~au) the jet axis, water masers trace three individual DW streamlines. By exploiting the 3D kinematic information of the masers, we determined the launch radii of these streamlines with an accuracy of \ $\sim$1~au, and they lie in the range of \ 10--50~au. At increasingly greater distances along the jet (110~au $\le xl \le 220$~au), the outflowing gas speeds up while it collimates close to the jet axis. Magneto-centrifugal launching in a radially extended MHD DW appears to be the only viable process to explain the fast (up to 60~\kms) and collimated (down to 10\degree) velocities of the wind in correspondence with launch radii ranging between 10~and~50~au. At larger separations from the jet axis ($R \ge 100$~au), the water masers trace a slow ($\le$20~\kms), radially expanding arched shock-front with kinematics inconsistent with magneto-centrifugal launching. Our resistive-magnetohydrodynamical simulations indicate that this shock-front could be driven by magnetic pressure.}
   {The results obtained in \targ\ demonstrate that VLBI observations of the 22~GHz water masers can reliably determine the launching mechanism of protostellar winds.}

\keywords{ISM: jets and outflows -- ISM: kinematics and dynamics -- Stars: formation -- Masers -- Techniques: interferometric}

\maketitle

\section{Introduction}
\label{Intro}

\begin{figure}
\includegraphics[width=0.5\textwidth]{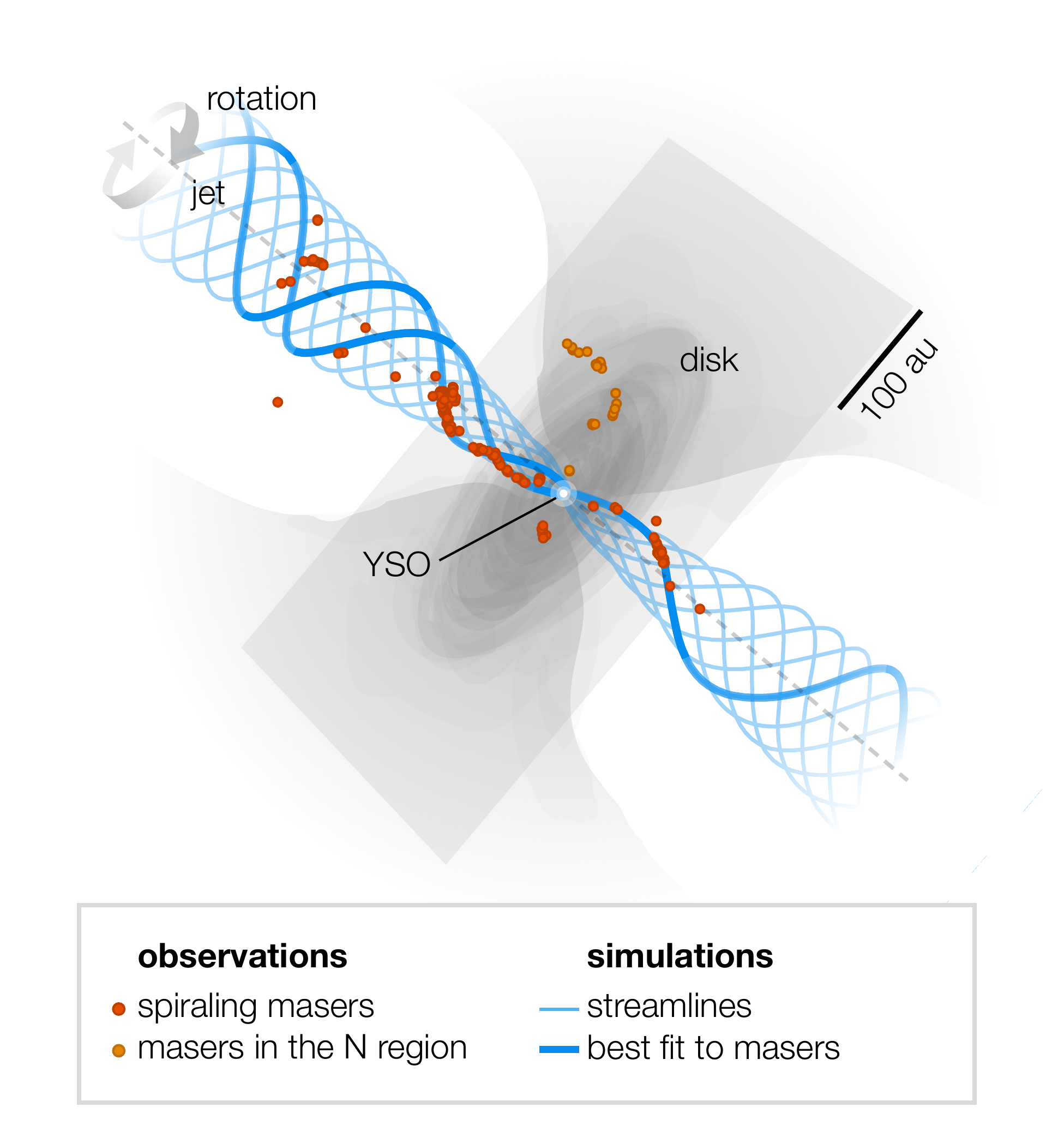}
\caption{{Three-dimensional view of the MHD DW in \targ\ from simulations in \citet{Mos23}.} 
The observed maser positions (red and orange dots) are overlaid on streamlines (blue lines) 
computed from resistive-radiative-gravito-MHD simulations of a jet around a forming massive star.}
\label{MHD-DW}
\end{figure}

Outflows in the form of either slow and poorly collimated winds or fast jets are commonly observed toward forming stars, which suggests the existence of a tight connection between mass accretion and ejection in the process of star formation. The ability of protostellar winds to extract angular momentum from the circumstellar disk and enable mass accretion, driving disk evolution and dispersal and ultimately setting the condition for planet formation, critically depends on their launching mechanism. Several mechanisms have been proposed to link accretion and ejection, and to account for the high degree of jet collimation, the magnetic field plays an important role in most of them. The proposed mechanisms include magneto-centrifugal acceleration either at the interface of the stellar magnetosphere with the disk \citep[X-winds,][]{Shu94} or from the centrifugally supported part of the disk \citep[disk wind, DW;][]{Bla82}, magnetic pressure alone \citep{Lyn03}, and thermal pressure following the disk photo-evaporation by intense stellar UV radiation \citep{Hol00}. 
Recent results on low-mass ($\sim$1~\ms) star formation favor models of magnetohydrodynamic (MHD) DWs since they can supersede magnetic turbulence and extract disk angular momentum in regions of low ionization and account for the emission of both winds and jets \citep{Pas23}. Protostellar jets have also been investigated in intermediate (a few solar masses) and high-mass ($\ge$10~\ms) protostars through high-angular resolution observations of the radio continuum and the measurement of proper motions of the radio knots \citep{Car21,Rod22}, with indications being found in favor of X-winds and DWs or jet collimation resulting from ambient gas pressure.

In the past ten years, high-angular resolution ($\approx$20--50~mas) Atacama Large Millimeter Array (ALMA) observations have identified several low-~and~high-mass young stellar objects (YSOs) where an LSR velocity (\Vlsr) gradient  transversal to the jet axis has clearly been detected. The latter is generally interpreted as jet rotation due to the magneto-centrifugal launching, and through various methods \citep{Tab20}, the launch radius of the wind has been derived. The launch radius, $R_{\rm K}$, is the critical parameter to distinguish between the quasi-polar X-winds \citep{Shu94}, $ R_{\rm K} \le 0.1$~au, from the radially extended DWs \citep{Pud07}, $ R_{\rm K} \sim$1--50~au, and this distinction is essential to understanding the wind's role in driving mass accretion and disk evolution. 
By targeting thermal lines, these ALMA observations reach linear resolutions of \ >10--50~au at best and can measure only line-of-sight velocities. 
Owing to the insufficient linear resolution and confusion between axial and rotation velocities, the ALMA measurements of the launch radii suffer from severe systematic errors and uncertainties. An emblematic case is that of the low-mass protostar HH212 \citep{Lee17}, which is probably the clearest example of jet rotation, where the observers estimated a launch radius for the SO wind of 1~au and an ad-hoc model reproducing the ALMA observations indicated a much larger value of 40~au \citep{Tab17,Tab20}. 

By achieving linear resolutions of $\sim$1~au and measuring 3D velocities, very long baseline interferometry (VLBI) observations of the 22~GHz water masers can directly trace the velocity field of  protostellar winds. Our previous studies \citep{Mos07,San10b,Mos11a,God11a} have demonstrated the ability of 22~GHz water masers in tracing the velocity structure of the outflows near YSOs. Recently, we carried out the Protostellar Outflows at the EarliesT Stages (POETS) survey \citep{Mos16,San18} to image the disk-outflow interface on scales of \ 10--100~au in a statistically significant sample (37) of luminous YSOs. We employed multi-frequency Jansky Very Large Array (JVLA) observations to determine the spatial structure of the ionized emission and multi-epoch Very Long Baseline Array (VLBA) observations (from the BeSSeL\footnote{The Bar and Spiral Structure Legacy (BeSSeL) survey is a VLBA key project whose main goal is to derive the structure and kinematics of the Milky Way by measuring accurate positions, distances (via trigonometric parallaxes), and proper motions of methanol and water masers in hundreds of high-mass star-forming regions distributed over the Galactic Disk \citep{Rei14}.} survey) to derive the 3D velocity distribution of the 22~GHz water masers. Among the main results of POETS, we find that the water masers are always associated with weak, slightly resolved continuum emission, tracing the thermal jet emitted from the YSO, and the 3D velocity distribution of the water masers in most of the sources of the sample can be interpreted in terms of an MHD DW. 

\targ \ is the most promising MHD DW candidate from the POETS survey.
The star-forming region has a bolometric luminosity of \ 5$\, \times \,$10$^{3}$~\ls\ \citep{Mos16} at a distance of \ 1.63$\pm$0.05~kpc \citep{Xu13} and harbors a cluster of forming massive stars. Around the most massive YSO ($5.6\pm2$~\ms) of the cluster, on scales of a few 100~au, a rotating disk has been observed in high-density molecular tracers by employing the NOrthern Extended Millimeter Array \citep[NOEMA;][]{Mos21}. The POETS JVLA 5~cm continuum observations have revealed a thermal jet \citep{Mos16} directed toward the NE-SW (PA $\approx$ 44\degree) and emerging from the YSO, whose position at the center of the disk is pinpointed by compact JVLA continuum at 1.3~cm. The BeSSeL VLBA observations of the 22~GHz water masers, performed in 2010--2011, have discovered a cluster of masers placed $\approx$100~au NE from the YSO and whose proper motions are collimated NE-SW (PA = 49\degree) and trace the base of the jet from the YSO. In October 2020, we re-observed the water masers in \targ\ with sensitive Global VLBI observations and found that the maser spatial distribution has significantly changed with respect to the previous BeSSeL epoch \citep{Mos22}. Remarkably, close to the YSO, the water masers trace sinusoids in the plane of the sky, which can be interpreted as the sky-projection of 3D helixes seen almost edge-on. The water masers could then be tracing the streamlines of an MHD DW, where the gas is predicted to flow along helical magnetic field lines anchored to the rotating disk. Indeed, \citet{Mos23} have shown that resistive-radiative-gravito-MHD simulations of a DW can model the maser streamlines with great accuracy (rms deviation of $\sim$1~au; see Fig.~\ref{MHD-DW}).

This article reports on new multi-epoch VLBA observations of the 22~GHz water masers in \targ carried out from March to October 2023. These new observations allow us to determine the 3D velocities of the water masers in the DW streamlines, which are key to accurately measuring the launch radii of the streamlines and reliably identifying the launching mechanism of the protostellar wind. Sect.~\ref{met_obs} describes the new VLBA observations, and the distribution of the water maser 3D velocities is presented in Sect.~\ref{Res}. The discussion of the new results is provided in Sect.~\ref{Dis} and the conclusions are drawn in Sect.~\ref{Conclu}.

\begin{figure*}%
\sidecaption
\includegraphics[width=12cm]{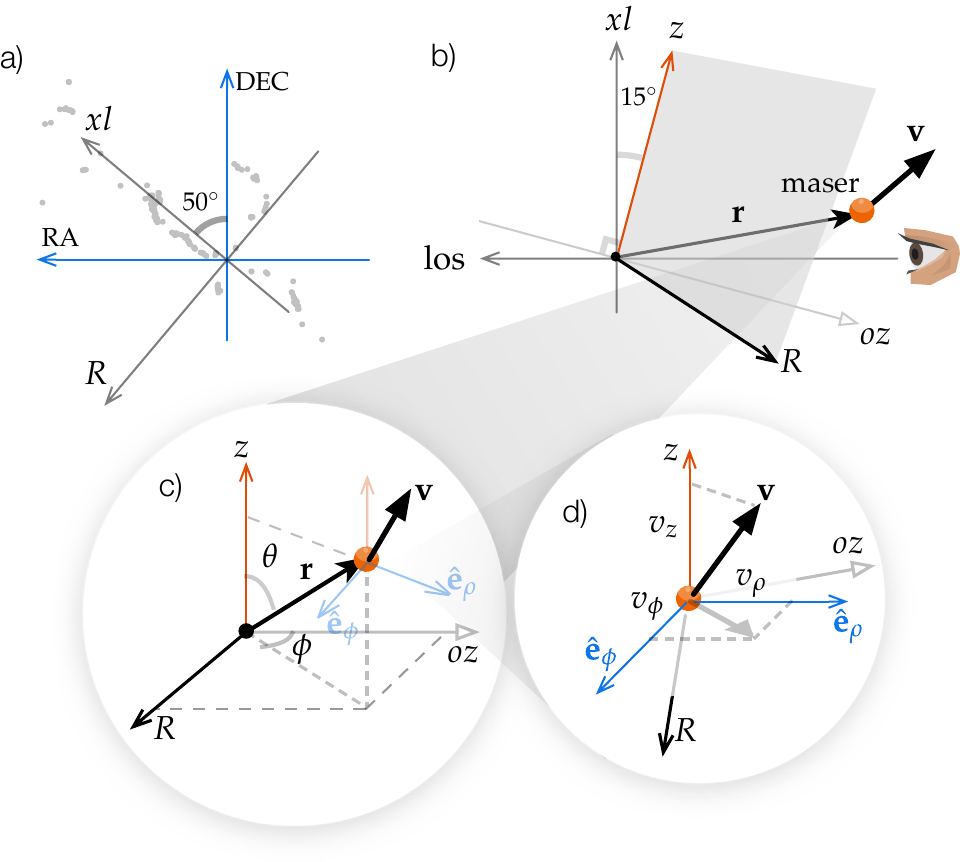}
\caption{{Reference frames used for maser positions and velocities, and magnetic fields.} \ (a)~Transformation of coordinates by rotation around the line-of-sight axis \ los \ over the position angle of the jet to bring \ DEC \ onto the sky-projected jet axis $xl$. \ (b)~Transformation of coordinates by rotation around the axis \ $R$ \ over the inclination angle of the jet to bring \ $xl$ \ onto the jet axis \ $z$. \ (c)~Transformation of coordinates by rotation around the axis \ $z$ \ over the rotation angle \ $\phi$ \ of the maser position to bring $oz$ along the direction \ $\hat{e}_{\rho}$. \ (d)~Decomposition of the maser velocity into the axial ($v_z$), radial ($v_{\rho}$), and azimuthal ($v_{\phi}$) components.}
\label{ax-tr}
\end{figure*}

\section{Very Long Baseline Array observations} 
\label{met_obs}

We observed the  $6_{16} - 5_{23}$ H$_2$O maser transition (rest frequency 22.235079 GHz) toward \targ\ 
(tracking center: RA(J2000) = $21^{\rm h} \, 9^{\rm m}$ 21\pss720 and Dec(J2000) = +52\degree\ $22^{\prime}$ 37\pas08) with the ten antennae of the  VLBA of the National Radio Astronomy Observatory (NRAO\footnote{NRAO is a facility of the National Science Foundation operated under cooperative agreement by Associated Universities, Inc.}) at four epochs: March 6, June 5, August 5, October 19, 2023. Each epoch lasted 14~hr. Because of technical problems, the antenna of Mauna Kea, Owens Valley, and Saint Croix respectively did not take part in the observations in the first, second, and fourth epochs. To determine the absolute positions and velocities, we performed phase-referencing observations, alternating scans on the target and the phase-reference calibrators every 45~s. The phase-reference calibrators were the quasars \ 2116$+$543 \ and \ 2051$+$528, both within 2.5\degree\ from the target and with a correlated flux  of $\sim$~0.1~\Jyb \ at 22~GHz. Calibration scans of $\approx$8~min were observed every hour on the fringe-finder and bandpass calibrators \ J2202$+$4216, 2007$+$777, 3C84, and 3C48.

We recorded dual circular polarization through four adjacent bandwidths of 16~MHz, one of them centered at the YSO's \Vlsr \ of \ -6.4~\kms. The four 16~MHz bandwidths were used to increase the signal to noise ratio of the weak continuum (phase-reference) calibrators. The data were correlated with the VLBA-DiFX software correlator at Socorro (New Mexico, USA) in two correlation passes using 1024 and 128 spectral channels to correlate the maser 16~MHz bandwidth and the whole set of four 16~MHz bandwidths, respectively. The spectral resolution attained across the maser 16~MHz band was 0.21~\kms. The correlator averaging time was 1~s. 

Data were reduced with the Astronomical Image Processing System \citep[\textsc{AIPS},][]{Gre03} package following the VLBI spectral line procedures in the \textsc{AIPS} COOKBOOK.\footnote{\url{http://www.aips.nrao.edu/cook.html}} At each epoch, the emission of an intense and compact maser channel was self-calibrated,
and the derived (amplitude and phase) corrections were applied to all the maser channels
before imaging. To cover the whole maser emission, we produced images extending  \ 1\pas23  \ in both \ RA $\cos \delta$ \ and \ DEC, and 84~\kms \ in \Vlsr. Reflecting the changes in the antennae taking part in the observations, using natural weighting, the FWHM major and minor sizes of the beam vary in the ranges \ 0.6--1.2~mas and 0.4--0.8~mas, respectively, and the beam PA in the range \  -3\degree--75\degree. In channel maps with a (relatively) weak signal, the 1$\sigma$ rms noise is \ $\approx$4~mJy~beam$^{-1}$, which is close to the expected thermal noise.

For a description of the criteria used to identify individual maser features, derive their parameters (position, intensity, flux, and size), and measure their (relative and absolute) proper motions, we refer to \citet{San10a}. Individual maser features are a collection of quasi-compact spots observed on contiguous channel maps and spatially overlapping (within their FWHM size). The spot positions are determined by fitting a two-dimensional elliptical Gaussian to their spatial emissions. The uncertainty of the spot position relative to the reference maser channel is the contribution of two terms: $\Delta \theta_{\rm spot} = \sqrt{\Delta \theta_{\rm fit}^2 + \Delta \theta_{\rm bandpass}^2}$. The first term depends on the S/N of the data, following \citet{Rei88}: \ $\Delta \theta_{\rm fit} = \theta_{\rm beam} \  / (2 \ {\rm S/N}) $, where \ $\theta_{\rm beam}$ \ is the resolution beam size, conservatively taken equal to the maximum value of the FWHM major beam size of 1.2~mas. The second term depends on the accuracy of the bandpass calibration through the expression \citep{Zha17}: \ $\Delta \theta_{\rm bandpass} = \theta_{\rm beam} \  (\Delta \Psi / 360^{\circ})$, where \ $ \Delta \Psi$ \ in degrees is the phase stability across the observing band. In our case, 
$ \Delta \Psi \lesssim 5^{\circ}$ \ and \ $\Delta \theta_{\rm bandpass} \lesssim 0.02$~mas becomes the dominant error term for spot intensity \ $\ge$~200~m\Jyb. The maser feature position (and corresponding error) is estimated from the error-weighted mean of the spot positions (and corresponding errors), and the feature \Vlsr \ from the intensity-weighted mean of the spots' \Vlsr. The error on the line-of-sight velocities, defined as the difference between the maser \Vlsr\ and the YSO's \Vlsr, is taken equal to the uncertainty of 0.4~\kms\ on the YSO's \Vlsr, exceeding the maser \Vlsr\ error of $\lesssim$~0.2~\kms.

Table~\ref{wat} reports the parameters (intensity, \Vlsr, position, and relative proper motion) of the 22~GHz water masers in \targ.
As described in \citet{San10a}, at least three observing epochs are necessary to assess the time persistence of a maser feature and accurately derive its proper motion. 
Relative proper motions are calculated with respect to the geometric center (hereafter ``center of motion", identified with label \#0 in Table~\ref{wat}) of the features with a stable spatial and spectral structure, persisting over the four observing epochs.
Errors of the relative proper motions are typically of \ $\approx$0.6~\kms.
At each epoch, inverse phase-referencing \citep{Rei09} produced good S/N ($\ge$~10) images of the two phase-reference calibrators. Taking into account that the calibrators are relatively compact, with size $\lesssim$~1~mas, and that the absolute position of the calibrators is known within a few 0.1~mas, we estimate that the error on the absolute position of the masers is \ $\lesssim$~0.5~mas. 
Without correction for the uncompensated delay introduced by the Earth's atmosphere \citep{Rei09}, our VLBA observations reach an astrometric accuracy of \ $\lesssim$~0.1~mas, and the absolute proper motions have typical errors of \ 2--3~\kms.  
The derived absolute proper motions have been corrected for the apparent proper motion due to the Earth's orbit around the Sun
(parallax), the solar motion, and the differential Galactic rotation between our LSR and that of the maser source. We adopted a flat Galaxy rotation curve ($R_0 = 8.33\pm0.16$~kpc, $\Theta_0 = 243\pm6$~\kms\ \citep{Rei14}) and the solar motion $U = 11.1^{+0.69}_{-0.75}$, $V = 12.24 ^{+0.47}_{-0.47}$, and $W = 7.25^{+0.37}_{-0.36}$~\kms\ by \citet{Sch10}, who revised the Hipparcos satellite results.

\section{Results}
\label{Res}

Figure~\ref{ax-tr} illustrates the reference frames employed to describe the maser positions and velocities, and magnetic fields. The jet axis $z$ is inclined 15\degree\ with the plane of the sky, and its projection on the sky \ $xl$ \ is directed at PA = 50\degree\ (see Appendix~\ref{met_ori}). The axis \ $R$ \ is orthogonal to  \ $xl$ \ in the plane of the sky, and the axis \ $oz$  \ is orthogonal to \ $z$ \ in the plane containing \ $z$ \ and the line-of-sight \ los. The maser positions in the reference frame \ ($R$, $oz$, $z$) \ can also be described using cylindrical coordinates: the inclination angle \ $\theta$ \ between the position vector \ $r$ \ and the axis \ $z$; $\rho$, the projection of \ $r$ \ into the plane \ $(R, oz)$, oriented at the azimuthal angle \ $\phi$ \ with respect to the axis \ $oz$. We note that \ $\phi$ \ increases clockwise in agreement with the disk rotation. Finally, velocity and magnetic field vectors can be decomposed in the reference frame co-moving with the masers along the axial ($\hat{e}_{z}$), radial ($\hat{e}_{\rho}$), and azimuthal ($\hat{e}_{\phi}$) directions.

\begin{figure*}%
\centering
\includegraphics[width=\textwidth]{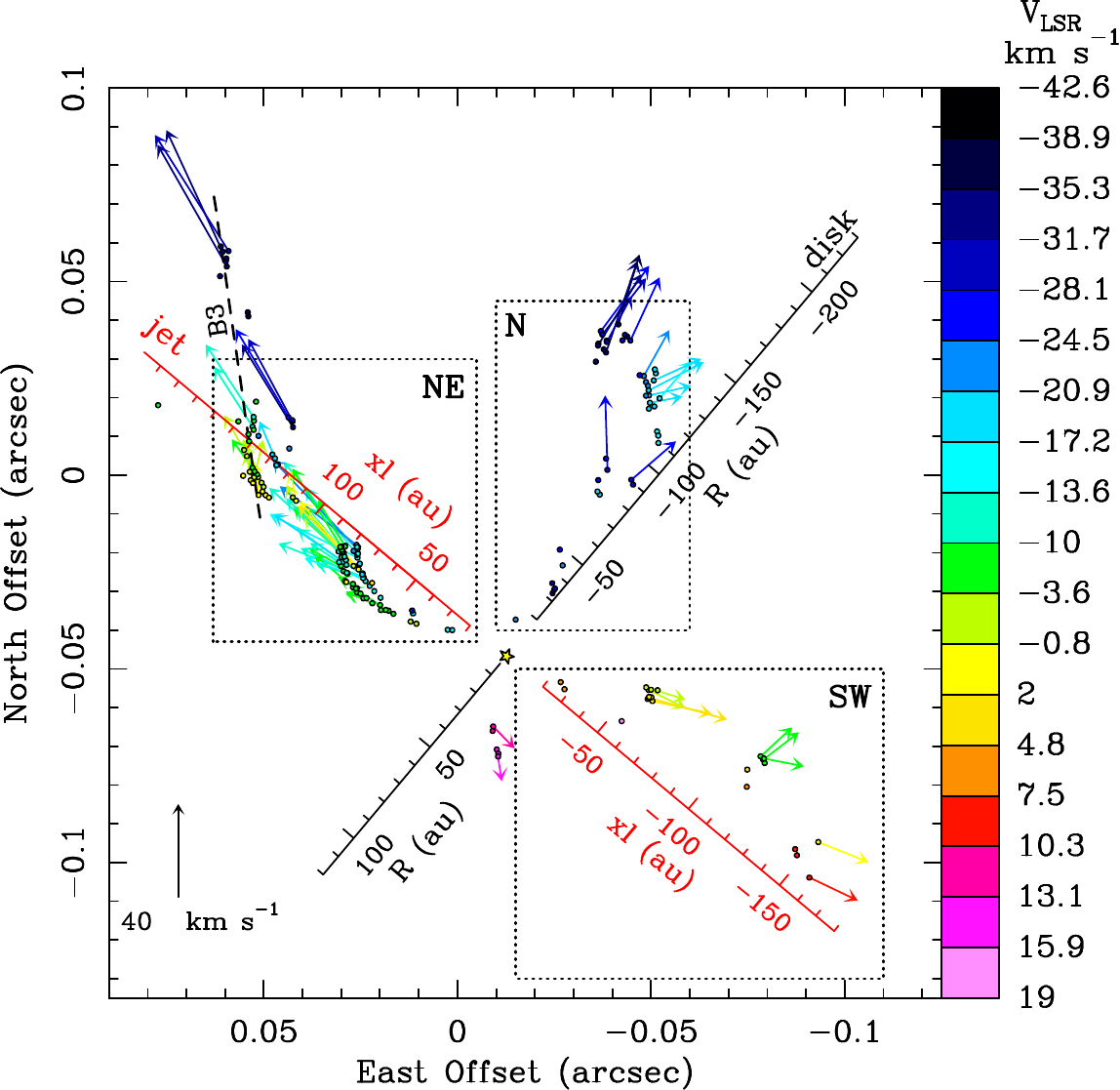}
\caption{{Water maser 3D velocities in \targ.} Colored dots and arrows give water maser positions (relative to the reference feature \#~0) and proper motions (relative to the YSO), respectively, with colors denoting the maser \Vlsr. We show only the masers for which the position at the first observing epoch can be derived. The amplitude scale for the velocity is reported in the bottom-left corner of the panel. The black dotted rectangles encompass the three regions to the N, NE, and SW of the YSO, where maser emission concentrates. The sky-projected jet and disk axes are in red and black, respectively, with distances labeled in astronomical units. An elongated maser stream directed approximately S--N is traced with a dashed-line segment labeled ``B3."}
\label{3V}
\end{figure*}

\begin{figure*}%
\includegraphics[width=0.49\textwidth]{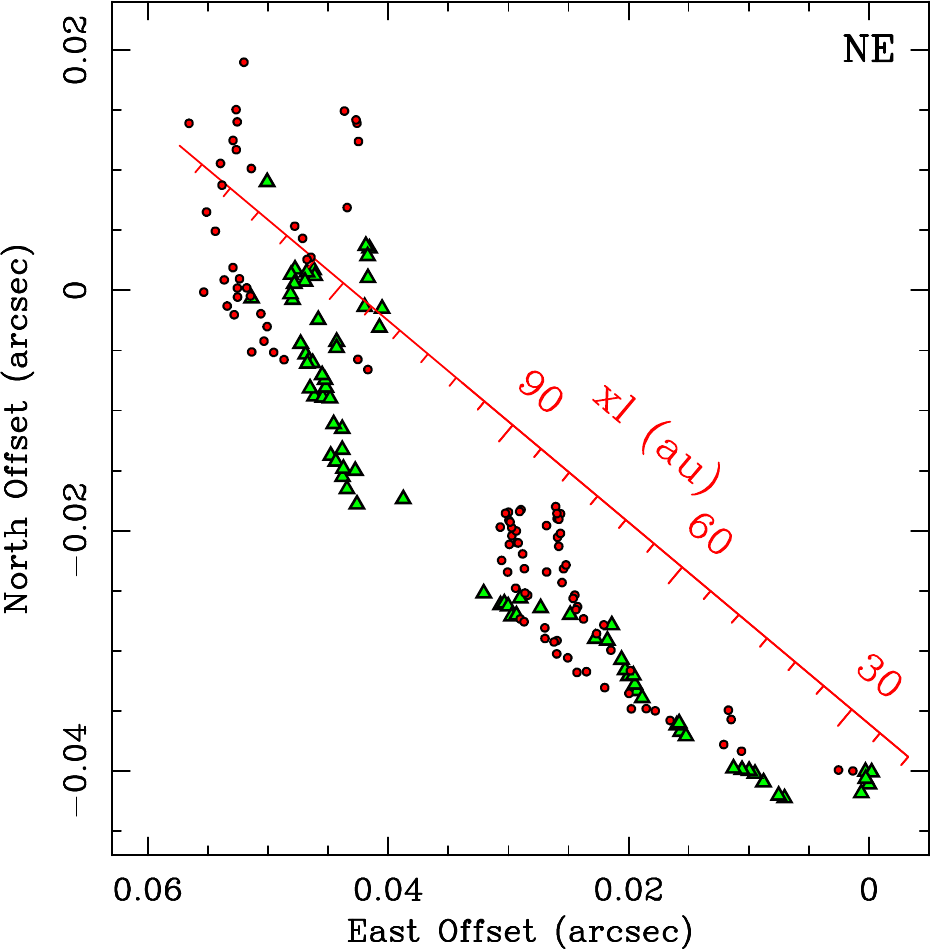}
\hspace*{0.2cm}\includegraphics[width=0.49\textwidth]{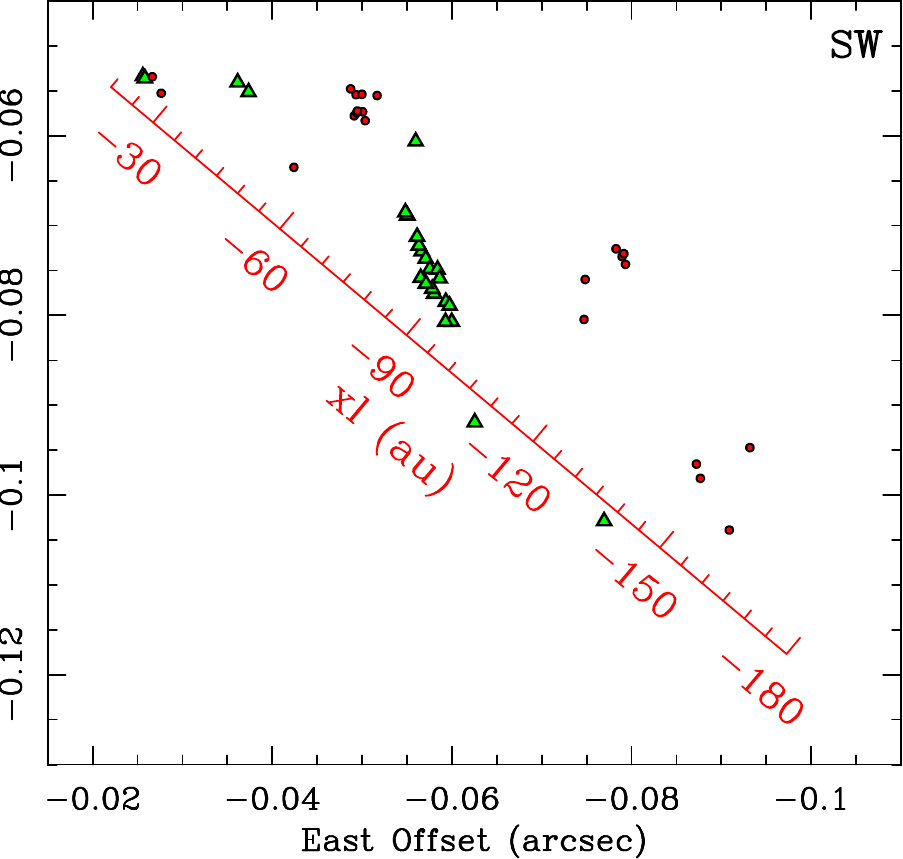}
\caption{{Time variability of the maser streamlines.} Comparison of the positions of water masers observed in October 2020 (green triangles) and March-October 2023 (red dots) for the NE (left) and SW (right) regions.}
\label{Fover}
\end{figure*}

\begin{figure*}%
\centering
\includegraphics[width=0.49\textwidth]{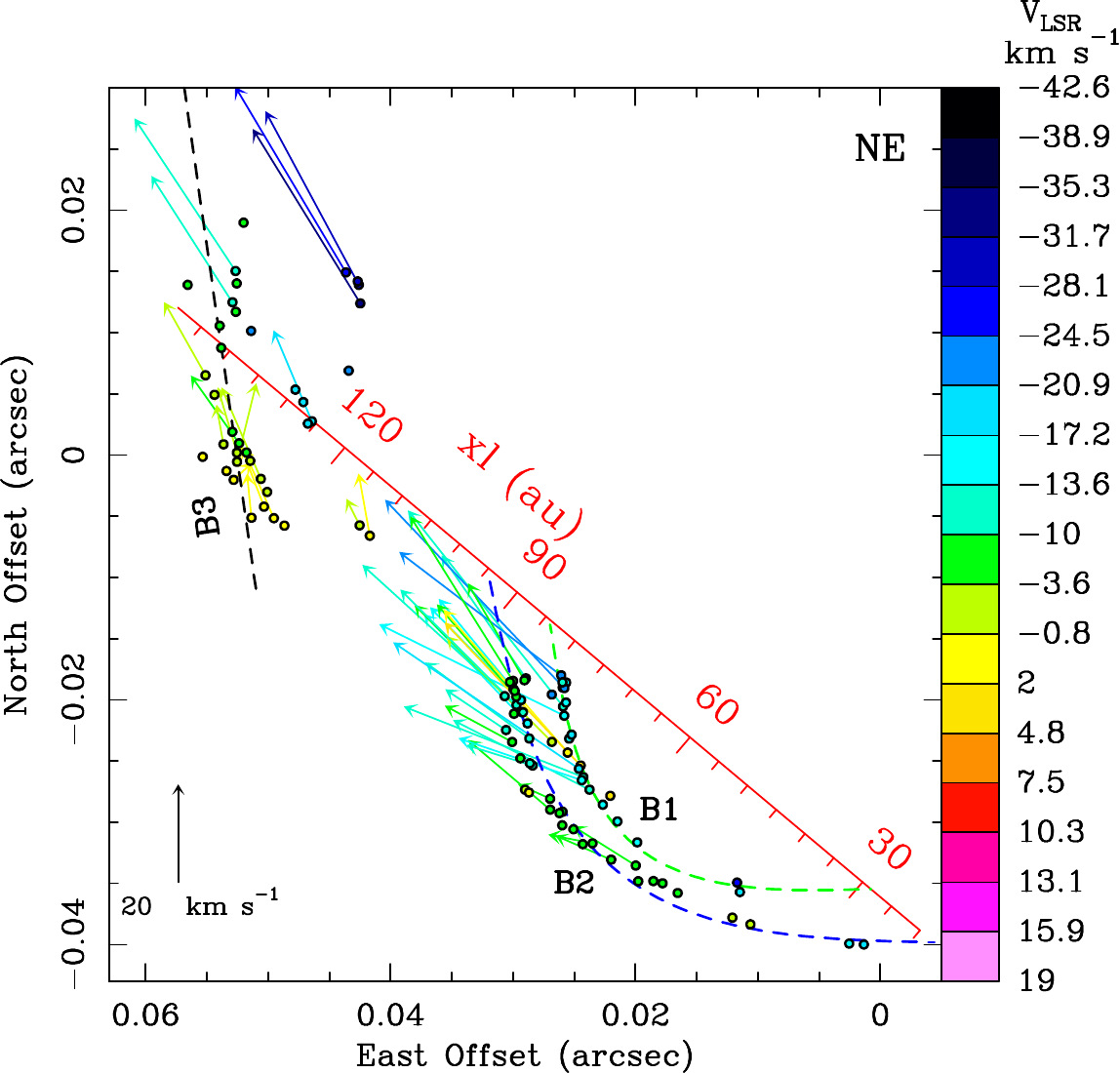}
\hspace*{0.15cm}\includegraphics[width=0.48\textwidth]{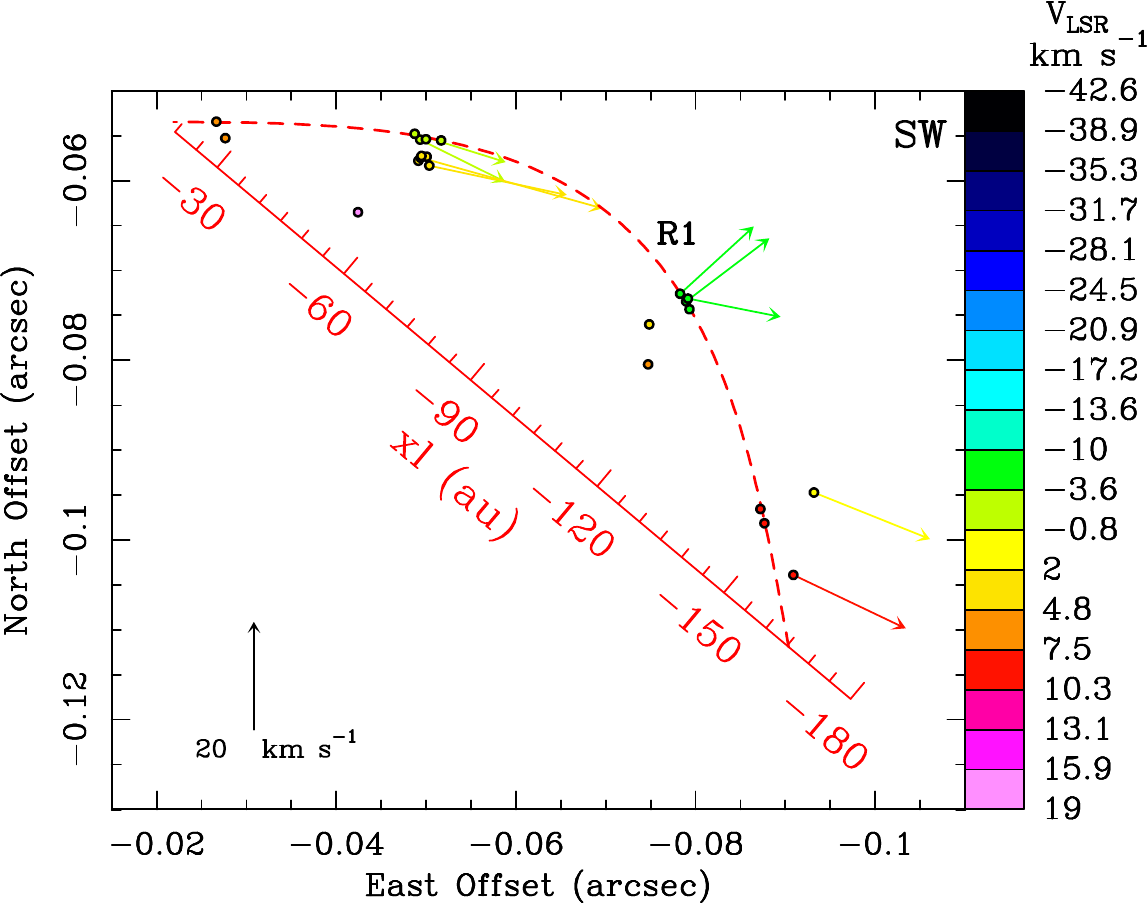}
\caption{{Three-dimensional velocities along the maser streamlines.} The colored dots and arrows and the red axes have the same meaning as in Fig.~\ref{3V}. (Left~panel)~Expanding view of maser positions and proper motions in the NE region. The green and blue dashed curves are the fitted sinusoids to the B1 and B2 streamlines, respectively, whose parameters are reported in Table~\ref{tab_fit}. \ (Right~panel)~Expanding view of maser positions and proper motions in the SW region. The red dashed curve is the fitted sinusoid to the R1 streamline, whose parameters are reported in Table~\ref{tab_fit}.}
\label{NE-SW}
\end{figure*}

Figure~\ref{3V} shows the maser positions and proper motions from the new 2023 VLBA observations. We note that the observing parameters of the 2023 VLBA observations, including beam size, spectral resolution, and sensitivity, are comparable to those of the previous October 2020 Global VLBI observations. The new observations continue to show a bipolar distribution of masers, both spatially and kinematically (in terms of proper motions and radial velocities), consistent with the findings of \citet{Mos22} from the 2020 observations. Maser emission is concentrated in the same three regions as in the previous Global VLBI observations, namely, to the NE, SW, and N of the YSO. However, inside each of these regions, the spatial distribution of the masers has changed. 
Figure~\ref{Fover} compares the positions of the water masers in the NE and SW regions between the 2020 Global VLBI and 2023 VLBA observations. In the NE region at \ 20~au~$\le xl \le$~90~au, the masers from 2020 traced a continuous curved pattern, while two separated streams at different distances from the jet axis are visible in the observations of 2023. In the SW region, the 2020 masers drew a single arc extending over \ $-100$~au~$\le xl \le$~$-20$~au; in 2023 there are hints of multiple arcs spreading over a larger distance of \ $-170$~au~$\le xl \le$~$-20$~au.

As in the 2020 observations, the maser patterns of the new 2023 VLBA observations still present sinusoidal shapes, which can be interpreted in terms of sky-projection of 3D helical streamlines.
Figure~\ref{NE-SW} shows a detailed view of these streamlines and the maser velocities along them. In the 2023 observations, inside the NE region over the $xl$ range 20--90~au, the masers trace two sinusoids of different amplitudes, named B1 and B2. In the SW region, only a single sinusoid, named R1, can be reliably identified with the masers. In Appendix~\ref{met_str} we discuss the procedure to fit the sinusoids to the maser positions, and the parameters of the sinusoidal fits ($\mathfrak{R}$, the amplitude of the sinusoid; $f_z$, the spatial frequency; and \ $xl_0$, the position of zero phase) are reported in Table~\ref{tab_fit}. A comparison of this table with Table~\ref{wat} in \citet{Mos22} confirms that the DW streamlines in \targ\ have changed over a timescale $\lesssim$1~yr.

\begin{table*}
\begin{flushleft}
\begin{minipage}{\textwidth}
\caption{{Parameters of the maser streamlines.}}
\label{tab_fit}
\begin{tabular*}{\textwidth}{@{\extracolsep{\fill}}lccccc@{\extracolsep{\fill}}}
\hline\hline
&  \multicolumn{3}{@{}c@{}}{Sinusoidal fit} \\ 
\cline{2-4}  
Stream & $\mathfrak{R}$ & $f_z$ & $xl_0$ &  $w_{\rm K}$ &  $R_{\rm K}$ \\
       & (au) & (rad~au$^{-1}$) & (au) &  (km~s$^{-1}$~au$^{-1}$) & (au) \\
\hline
B1  &  $15.7\pm0.3$ &  $0.047\pm0.001$ & $28.5\pm0.9$ &  $2.17\pm0.23$ & $10.8\pm0.8$ \\
B2  &  $19.9\pm0.4$ &  $0.034\pm0.001$ & $17.5\pm0.9$ &  $0.91\pm0.20$ & $19.2\pm2.8$ \\
R1  &  $39.3\pm0.2$ &  $0.018\pm0.001$ & $-165.4\pm0.5$ &  $0.22\pm0.03$ & $49\pm5$ \\
\hline
\end{tabular*}
\footnotetext{
\\
Column~1 denotes the maser stream.
Columns~2,~3,~and~4 report the amplitude, the spatial frequency, and the position of the zero phase, respectively, of the sinusoidal fit of the maser coordinates \ $R$ \ versus \ $xl$. The spatial frequency $f_z$ has been corrected by projection effects caused by the inclination of the jet axis ($i_{\rm jet} = 15$\degree) using the expression $ f_{z} =  \pi / (((\pi / f_{xl}) + 2 \, \mathfrak{R} \sin(i_{\rm jet})) / \cos(i_{\rm jet}))  $, where $f_{xl}$ is the original value from the sinusoidal fit. 
Columns~5~and~6 list the derived Keplerian angular velocity and launch radius. 
}   
\end{minipage}
\end{flushleft}
\end{table*}

\begin{SCfigure}
\includegraphics[width=0.35\textwidth]{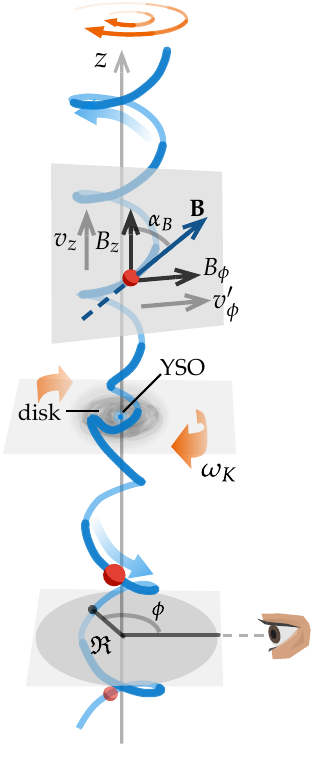}
\caption{{Co-rotating helical magnetic field line.} The water masers (red balls) move along a helical magnetic field line (blue) anchored to the rotating disk. The rotation radius and angle, $\mathfrak{R}$ and $\phi$, and the axial, $B_{z}$ \ and \ $v_{z}$, and azimuthal, $B_{\phi}$ \ and \ $v_{\phi}^{\prime}$, components of the magnetic field and velocity in the reference frame co-rotating with the disk are labeled. The tangent to the helix is inclined by an angle $\alpha_{\rm B}$ with respect to the $z$ axis.}
\label{helix}
\end{SCfigure}

As discussed in Appendix~\ref{met_ref}, the proper motions presented in Fig.~\ref{3V} are calculated relative to the YSO. The uncertainty is of a few kilometers per second and velocity amplitudes range from a few kilometers per second up to $\approx$60~\kms, with the fastest motions found close to the jet axis at high $xl$ $\ge$~120~au. Some of the masers with the highest velocities belong to the B3 stream, an elongated distribution of masers that extend from the NE region northward. We note that the spatial distribution of the masers of the B3 stream inside the NE region (over \ 105~au~$\le xl \le$~150~au; see Fig.~\ref{NE-SW}, left~panel) is not actually linear but slightly arched. However, as discussed in Appendix~\ref{met_str}, the positional scatter of these masers is too large to permit a reliable determination of a streamline. The masers in the N region trace an extended ($\approx$50~au) arc (see Fig.~\ref{3V}), with proper motions directed about perpendicular to the arc, suggesting that it is expanding away.

\section{Discussion}
\label{Dis}

\subsection{The magneto-centrifugal disk wind}
\label{Dis_DW}

Figure~\ref{helix} presents a sketch of an MHD DW streamline where the gas flows along a helical magnetic field line anchored to the rotating disk.
According to the MHD DW theory, the gas is magneto-centrifugally launched along a quasi-poloidal magnetic field line departing from the disk. The field line behaves like a rigid flow tube co-rotating with the disk until reaching the Alfv\'en point where the increasing kinetic energy of the outflowing gas equals the magnetic energy. Beyond that point, the inertia of the flow becomes important, the magnetic field line starts lagging behind the disk rotation, and the azimuthal component of the magnetic field grows in the rotation verse opposite to the disk. Thus, the geometry of the magnetic field line approaches that of a helix winding around the disk axis in counter-rotation with the disk.

The shape of the helical trajectories of the masers is characterized by the magnetic field helix angle \ $\alpha_B = \arctan(B_{\phi} / B_{z}) $, which is the angle with which the helical magnetic field winds around the jet axis (see Fig.~\ref{helix}). The terms $B_{z}$ \ and \  $B_{\phi}$ \ are the axial and azimuthal components of the magnetic field, respectively. 
In particular, by performing a sinusoidal fit of the sky-projected maser streamlines, one can derive \ $\mathfrak{R}$, the amplitude of the sinusoid, corresponding to the radius of the helix and  $f_z$, the spatial frequency of the sinusoid (see Sect.~\ref{Res} and Table~\ref{tab_fit}).
Following Eq.~12 in \citet{Mos22}, \ $\| B_{\phi} \| / \| B_{z} \| = \|  \tan(\alpha_B)  \| = f_z \ \mathfrak{R} $, that is, the product $ f_z \ \mathfrak{R}$ \ defines the shape of the helix. In the above equations, we have accounted for the negative sign of \ $B_{\phi}$, \ as it is directed opposite to the verse of increasing $\phi$.

An observer co-rotating with the launch point of the streamline (at the angular velocity $\omega_{\rm K}$) sees the gas moving along the helical field line. That is, in such a reference frame, the flow velocity is parallel to the magnetic field. Therefore, we can write
\begin{equation}
\frac{\| v_{\phi}^{\prime} \|}{\| v_{z} \|} = \frac{\| B_{\phi} \|}{\| B_{\rm z} \|} = f_z \ \mathfrak{R} \label{VpB},
\end{equation}
where we use the apical prime to indicate that the azimuthal velocity is measured in the co-rotating reference frame. On the other hand, the maser motion we observe is the composition of the motion along the streamline and the rotation of the streamline anchored to the disk. While the axial component of the velocity is not affected by such a rotation, the azimuthal velocity component we measure is given by
 \begin{equation}
v_{\phi} = v_{\phi}^{\prime} + \omega_{\rm K} \, \mathfrak{R} \label{vphi}.
 \end{equation}

Combining Eqs.~\ref{VpB}~and~\ref{vphi}, we derive
\begin{equation}
\omega_{\rm K}  = \frac{v_{\phi}}{\mathfrak{R}} + f_z \, \| v_{z} \|   \label{wK},
\end{equation}
which shows that the knowledge of the streamline shape and maser 3D velocity allows us a direct determination of the angular velocity of the launch radius.
Our previous Global VLBI observations \citep{Mos22} of \targ\ measured just the maser line-of-sight velocity, and we could only set lower limits for the axial velocities of the masers and the angular velocities of the launch radii of the streamlines and only derive upper limits for the launch radii.

\begin{table}
\begin{minipage}{0.5\textwidth}
\caption{{Three-dimensional velocity in the streams.}}
\label{tab_vel}
\begin{tabular}{cccc} 
\hline\hline
Stream &  $v_{\rm z}$  &  $v_{\rm \rho}$ &  $v_{\rm \phi}$  \\
       &  (km~s$^{-1}$) & (km~s$^{-1}$) & (km~s$^{-1}$) \\
\hline
B1  &  $42\pm9$ & $2\pm6$ & $3\pm6$ \\
B2  &  $30\pm6$ & $-5\pm4$ & $-2\pm4$ \\
R1  &  $-11\pm6$ & $10\pm6$ & $3\pm5$ \\
N   &  $7\pm10$ & $32\pm8$ & $2\pm4$ \\
\hline
\end{tabular}
\footnotetext{
\\
Column~1 denotes the maser stream; 
Columns~2,~3,~and~4 report the values of $M\pm\sigma$, where $M$ is the mean and  \ $\sigma$ \ is the standard deviation, for the velocity components along the jet axis, radial, and azimuthal, respectively.
The azimuthal angle of the masers in the N region is taken equal to the average azimuthal angle, 304\degree, of the velocities.
}   
\end{minipage}
\end{table}

Knowing the jet PA = 50\degree\ and its inclination \ $i_{\rm jet} = 15$\degree\ with the plane of the sky  (see Appendix~\ref{met_ori}), we can convert the velocity components from the reference system  \ (RA, DEC, los) \ in which they are originally measured to the reference system \ ($R$, $oz$, $z$; see Figs.~\ref{ax-tr}a~and~\ref{ax-tr}b). The rotation (or azimuthal) angle, $\phi$, of the masers belonging to a streamline is calculated with the formula: \ $ \phi = \arcsin(R / \mathfrak{R})$, where $R$ is the sky-projected radius of the maser and \ $\mathfrak{R}$ \ is the rotation radius of the streamline. By fitting sinusoids to the sky-projection of helical motions and varying the inclination angle of the helix axis, we verified that this formula is accurate to within a few degrees for small inclinations \ $\le 20$\degree. As the streamline winds counterclockwise around the disk axis, the rotation angle goes from 180\degree\ to 0\degree\ for increasing \ $xl$ \ along the two NE streamlines, and from 360\degree\ to 180\degree\ for decreasing \ $xl$ \ along the SW streamline. The knowledge of the rotation angle of a given maser allows us to determine the radial, $v_{\rho}$, and azimuthal, $v_{\phi}$, components of its velocity  (see Figs.~\ref{ax-tr}c~and~\ref{ax-tr}d). Table~\ref{tab_vel} reports the mean and standard deviation of the axial, radial, and azimuthal velocity components for each streamline. While \ $v_z$ \ dominates over $v_{\rho}$ and  $v_{\phi}$ \ in the B1 and B2 streamlines,  $v_{\rho}$ and $v_z$ \ are comparable in the R1 streamline.

\begin{figure*}
\centering
\includegraphics[width=0.67\textwidth]{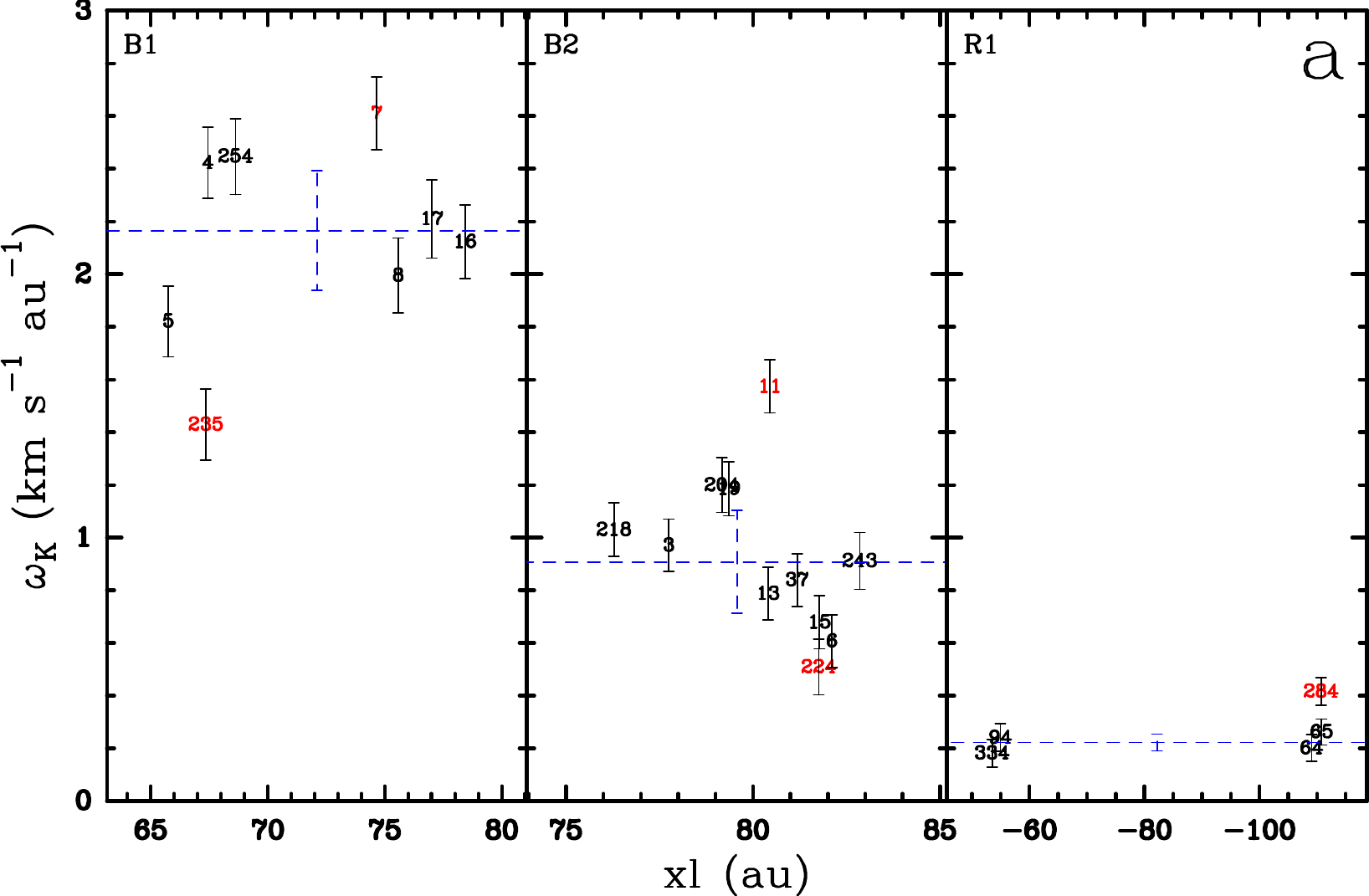}
\hspace*{0.0cm}\includegraphics[width=0.32\textwidth]{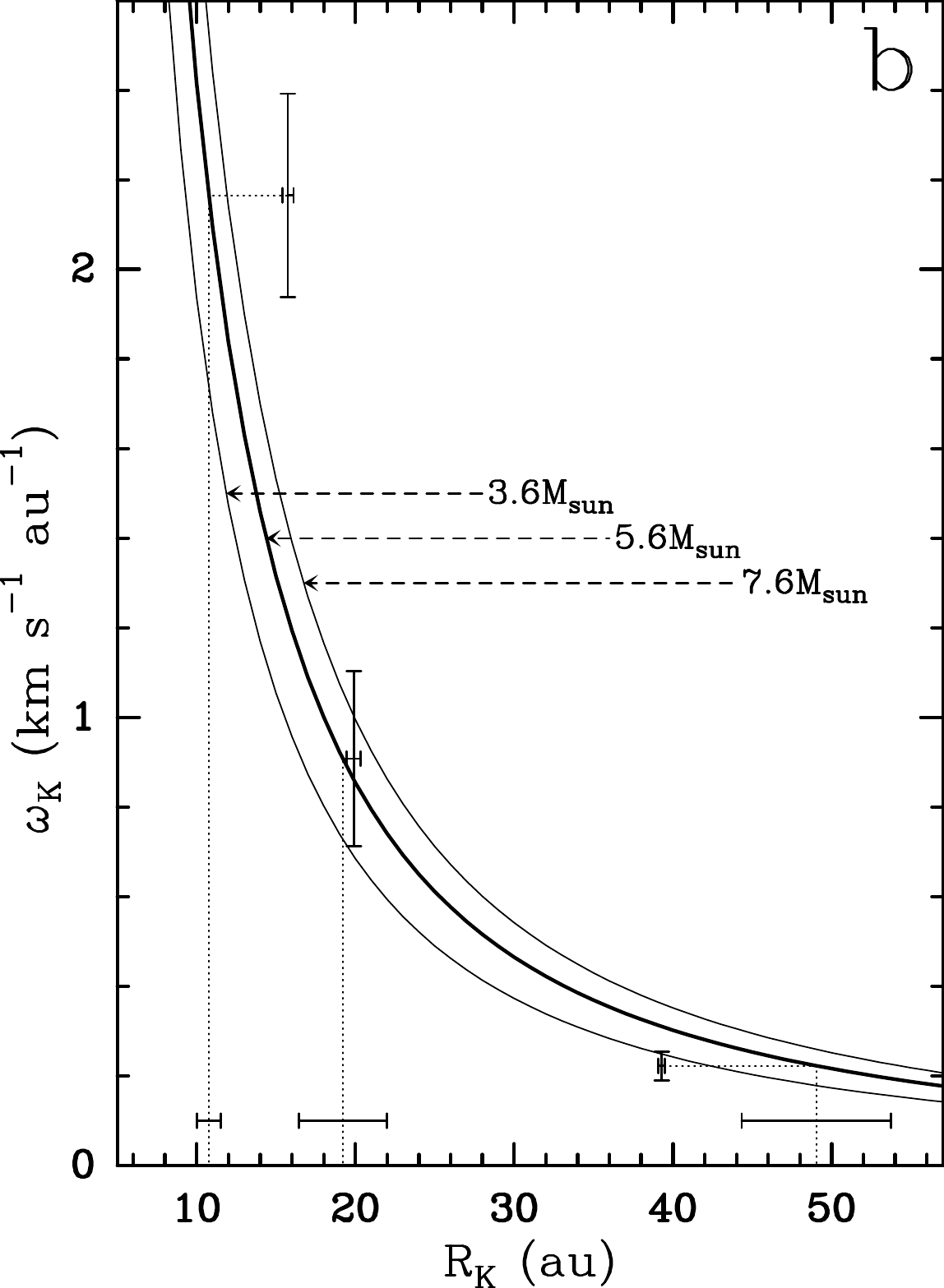}
\caption{{Determination of the launch radii of the maser streamlines.} \ (a)~The streamline to which each panel refers is indicated in the top-left corner of the panel. Maser labels (see Table~\ref{wat}) and error bars indicate the values (and corresponding errors) of the angular velocity of the launch radius inferred with the masers belonging to the streamline, which are plotted versus the position along the sky-projected jet axis \ $xl$. The dashed blue line and error bar give the mean and standard deviation of the consistent values, excluding the outliers denoted with red labels. \ (b)~Vertical and horizontal error bars plot the values (and associated errors) of the angular velocity and rotation radius of the streamlines. The heavy and thin curves are the Keplerian profiles evaluated at three different YSO masses, as labeled. The launch radius of each streamline for the nominal value of the YSO mass of \ 5.6~\ms\ is identified with horizontal and vertical dotted lines, and the associated measurement error is marked by the horizontal error bar in the bottom.}
\label{FwK}
\end{figure*}

Figure~\ref{FwK}a shows the distribution of the Keplerian angular velocities computed from all masers using Eq.~\ref{wK}. In agreement with our expectation, the large majority (75-80\%) of the masers belonging to the same streamline provide consistent values of the angular velocity. We emphasize that the found consistency provides a direct test of the MHD DW model employed to interpret the data. A low fraction of outliers\footnote{We determined the outliers by calculating the average separation AS of each value from the others, evaluating the mean \ $M_{\rm AS}$ \ and standard deviation \ $\delta_{\rm AS}$ \ of the AS, and considering those values whose \ AS $ > M_{\rm AS} + \delta_{\rm AS}$.} can be expected considering both the high time variability and the possibility of spatial overlap of the streamlines (see Fig.~\ref{Fover} and Appendix~\ref{met_str}). Our method of filtering out the outliers is mainly based on the scattering of the maser positions from the sinusoidal fit and can be ineffective at the loci of intersection of two different streamlines. 

Table~\ref{tab_fit} lists the value (and corresponding error) of the angular velocity of the launch radius for each streamline, calculated by taking the mean (and standard deviation) of the consistent values of angular velocities in Fig.~\ref{FwK}a. By reporting the derived angular velocities on the Keplerian profile of the YSO, Figure~\ref{FwK}b shows that the launch radii of the three streamlines can be determined with high accuracy:  10.8$\pm$0.8~au,  19.2$\pm$2.8~au, and  49$\pm$5~au for B1,  B2, and R1, respectively. If the uncertainty of 2~\ms \ on the YSO mass is taken into account, the values of the launch radii change by \ $\approx$12\%. We stress that our findings are crucial to favor the DW model over the X-wind model, as the latter predicts launch radii $< 0.1$~au. Besides, the high velocities (up to 60~\kms) observed close to the jet axis (see Fig.~\ref{3V}) are inconsistent with acceleration due to (only) magnetic or thermal pressure, which produce typical terminal flow speeds  $< 20$~\kms\ \citep{OK23b}. In conclusion, magneto-centrifugal launching in a radially extended MHD DW appears to be the only viable process to explain the maser 3D velocity field in this source. 

%

\subsection{Magneto-centrifugal launching versus magnetic pressure}
\label{Dis_MP}

\begin{figure*}%
\sidecaption
\includegraphics[width=12cm]{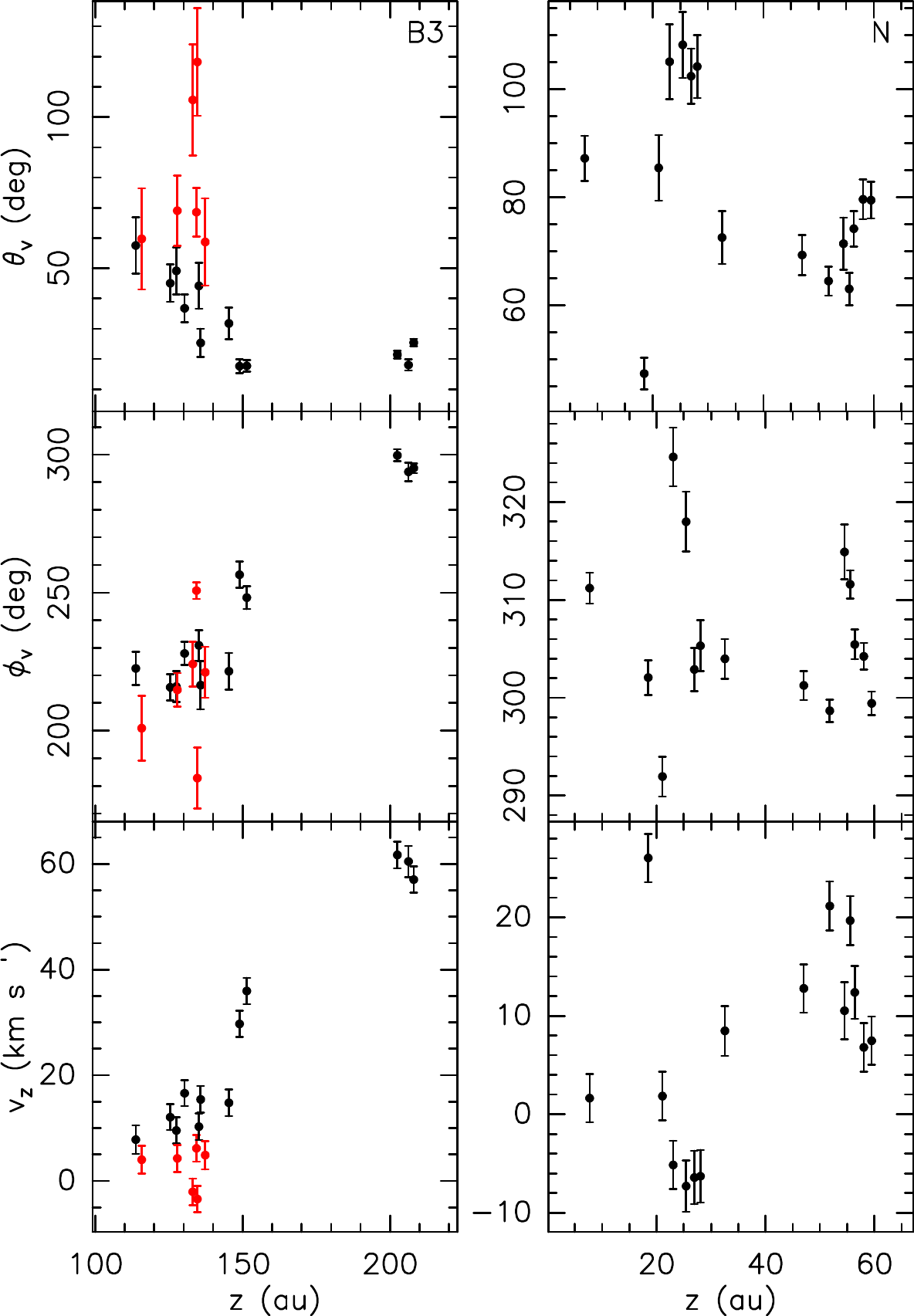}
\caption{{Change in the maser velocity direction and amplitude moving along the jet axis.} The left and right panels refer to the B3 stream and N region, respectively. The inclination, $\theta$, and \ azimuthal, $\phi$, angles of the direction of the velocity (see Fig.~\ref{ax-tr}c)\  and\ the velocity component along the jet, $v_z$, are plotted versus the position along the jet, $z$, \ in the top, middle, and bottom panels, respectively. Values and corresponding errors are denoted with dots and error bars, respectively, where red marks velocities \ $\lesssim$6~\kms\ in the B3 stream.}
\label{NV}
\end{figure*}

Figure~\ref{NV} compares the velocities of the masers in the B3 stream and N region. In the B3 stream, for increasing \ $z$, the velocities of the masers grow and collimate toward the jet axis, from $v_z \le $10~\kms\ and \ $\theta_V \ge$ 60\degree\ at $z \approx 100$~au to  $v_z \approx$60~\kms\  and \ $\theta_V \approx$10\degree\ at $z \ge 200$~au. The large change of $\approx$100\degree\ in the azimuthal angle $\phi_V$ can be interpreted in terms of rotation of the velocity vector around the jet axis. The acceleration to large ($\gtrsim 100$~\kms) speeds together with the collimation of the velocities close to the jet axis represent a distinct characteristic of the magneto-centrifugal launching of an MHD DW \citep{Pud07}. Therefore, the analysis of the maser kinematics in the B3 stream and in the B1, B2, and R1 streamlines leads us to conclude that all the outflowing gas close ($R \le 50$~au) to the jet axis is magneto-centrifugally launched by an MHD DW. As previously investigated in simulations \citep{OK23b}, the acceleration of the flow occurs when the magnetic field guides the flow outward, where the gravitational energy is less negative. The flow then becomes super-Keplerian and runs away: the kinetic energy keeps increasing along the magnetic field line, carrying angular momentum outward in the process. The kinematics of the masers in the N region, at $R \ge 90$~au, is remarkably different from that in the B3 stream. In the N region, masers are found at relatively low heights, $z \le$60~au, and move at small speeds, $\le$~20~\kms, along directions about parallel to the disk (70\degree\ $\lesssim \theta_V  \lesssim$~110\degree). The arc-like spatial distribution (see Fig.~\ref{3V}) together with the small spread in the azimuthal angle of the velocities, $\phi_V \approx$~310\degree$\pm$10\degree\ (see Fig.~\ref{NV}), suggest that these masers trace a radially expanding narrow shock-front placed at relatively large radii, 90~$\le R \le$~130~au (see Fig.~\ref{3V}), and extending from the disk plane up to $z \approx$60~au. This kinematic pattern is clearly inconsistent with magneto-centrifugal launching, and instead it could be explained in terms of acceleration by magnetic pressure, which, according to our MHD DW model \citep{Mos23}, dominates at $R \ge$~50~au.

\begin{figure*}
        \centering
        \includegraphics[width=\textwidth]{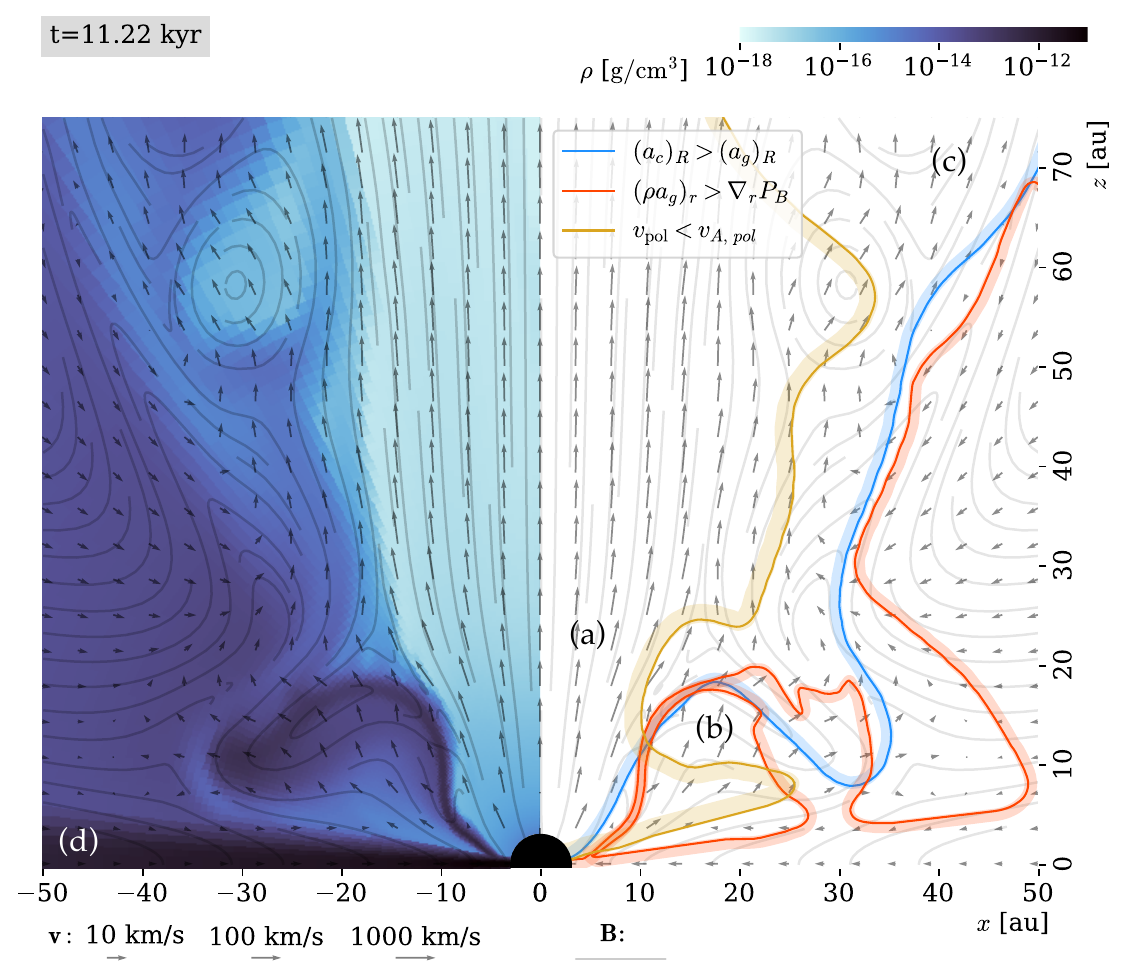}
        \caption{Snapshot of the simulation when the mass of the protostar is $5\,M_\odot$. The left panel shows the density, velocity, and magnetic field lines in the circumstellar gas.  The right panel contains contours (solid lines) accompanied by shadows (semi-transparent thick lines of the same color as the contour). The shadow indicates in which direction the inequality listed in the legend holds true. A smoothing algorithm has been applied to the contours to improve clarity. For more details, we refer to \citet{OK23b}.}
        \label{MP}
\end{figure*}

We previously presented \citep{Mos23} a resistive-magnetohydrodynamical simulation (including radiation transport and self-gravity) that closely (rms deviation of \ $\sim$1~au) reproduces the maser spatial patterns of the 2020 Global VLBI observations (see Fig.~\ref{MHD-DW}). We looked for simulations that simultaneously reproduce the mass of the protostar, the disk radius, the presence of a high velocity jet, and the accretion rate observed in \targ. We then explored several launch radii for the velocity streamlines (in the frame of reference co-rotating with the launch point) and compared against the maser spatial patterns. In the simulations, we previously found an Alfv\'en surface $\lesssim 50\,\mathrm{au}$ wide that becomes more conical toward the protostar because the outflow is constrained by the more dense material from the disk and infalling envelope. The arc-like maser distribution in the N region is located at $R \gtrsim 100\,\mathrm{au}$ away from the protostar. Even though we cannot rule out that the magneto-centrifugal mechanism can operate at such distances (see discussion on the caveats of the simulations in our previous articles \citep{Mos23,OK23a}), the condition that the Alfv\'en speed in the launching region be sufficiently large must be satisfied. This means that the density must be sufficiently low and the magnetic field sufficiently strong. Both conditions are less likely to be satisfied at large distances from the protostar as the infalling material is encountered.

On top of the accretion disk, the rotational dragging of the magnetic field lines create a magnetic pressure gradient that can be stronger than gravity (see our previous discussion in \citet{OK23b} on magnetic tower flows). Given that the maser kinematics in the N region do not exhibit the signatures of magneto-centrifugal motion, we examined whether magnetic pressure could be the driving force behind their kinematics. Figure \ref{MP} shows a snapshot of the simulation including an analysis of forces (right panel). The label \textit{a} in the figure indicates the magneto-centrifugal jet, which is launched and accelerated in the region where the flow is simultaneously sub-Alfv\'enic (shadowed yellow contour) and centrifugal force dominates over gravity (blue shadowed contour). The region where the magnetic pressure gradient dominates over gravity is bigger than the Alfv\'en surface (cf. unshadowed part of the red contour). As a consequence, regions \textit{b} and \textit{c} are outflowing but not magneto-centrifugally launched. Region \textit{b} is an episodic phenomenon we previously described as a cavity wall ejection \citep{OK23b}. At this stage, the radially outward motion in \textit{b} is driven by magnetic pressure, which is accumulated on top of the accretion disk (\textit{d}) at a similar height (but not radial distance) as the masers in the N region. Additionally, most of the material in \textit{b} is super-Alfv\'enic, meaning that we expect no correlation of the magnetic field direction and flow, just as observed in the polarization data from the N region \citep{Mos23}. A magnetic tower flow (\textit{c}) should have an important vertical velocity component. Because the proper motions in the N region exhibit a rather weak $v_z$, we conclude that a plausible production mechanism for the kinematical features of the masers in the N region is an episodic ejection event. In the simulations, such episodic ejection phenomena have lifetimes of 10--100 years. Future observations will provide a better understanding of the nature of the flow in the N region.

\section{Conclusions}
\label{Conclu}

We have recently carried out the POETS survey, where the 3D velocities of the protostellar winds in a relatively large (37) sample of luminous YSOs have been mapped at linear resolutions of $\sim$1~au through VLBI observations of the 22~GHz water masers. This technique overcomes the limitations of interferometric observations of thermal tracers, which, being significantly more extended than the maser cloudlets, are less suitable to act as kinematic probes of the small regions where protostellar winds are launched and collimated. The POETS results suggest that MHD DWs can be the common launching mechanism of protostellar winds in high-mass YSOs. In this article, we have tested the MHD DW scenario in the YSO \targ, the best MHD DW candidate from the POETS survey.
We have performed multi-epoch VLBA observations of the 22~GHz water masers in \targ\ to measure the 3D velocities of the maser cloudlets tracing the DW streamlines.

In \targ, the fast (10--60~\kms) outflowing gas moving at small angles ($\lesssim$~30\degree)  and separation (R $\lesssim$40~au) from the jet axis is magneto-centrifugally launched in a radially extended MHD DW. The water masers trace three DW streamlines whose launch radii are measured with an accuracy of \ $\sim$1~au: 10.8$\pm$0.8~au,  19.2$\pm$2.8~au, and  49$\pm$5~au. We have directly tested the model of an MHD DW by verifying that the water masers move along individual DW streamlines co-rotating with the disk at the angular velocity of their launch radius. At larger radii (R $\gtrsim$100~au), the relatively slow ($\lesssim$~20~\kms) flow traced with the masers propagates outward mainly along the radial direction, and its properties are clearly inconsistent with magneto-centrifugal acceleration. The comparison with our resistive-magnetohydrodynamical simulations suggests that these masers can trace an episodic ejection event driven by the magnetic pressure accumulated on top of the accretion disk. This work proves that water maser VLBI observations can determine the launching mechanism of protostellar winds in a trustworthy manner, as the model assumptions can also be tested, and it calls for applying this technique across the range of protostellar masses and ages traceable with water masers. Our findings are in line with recent results of low-mass star formation and suggest that MHD DWs could be the common launching mechanism of protostellar winds from low to high-mass protostars.

\begin{acknowledgements}
  AO acknowledges financial support from the Federal Commission for Scholarships for Foreign Students for the Swiss Government Excellence Scholarship (ESKAS No. 2023.0405) for the academic year 2023-2024, the Office of International Affairs and External Cooperation of the University of Costa Rica, and the European Research Council under the European Union's Horizon 2020 research and innovation program (project ID 833925, project STAREX).  
O.B. acknowledges financial support from the Italian Ministry of University and Research - Project Proposal CIR01\_00010.
 Scientific results from data presented in this publication are derived from the following VLBA (NRAO) project code:  BM534.
NRAO is a facility of the National Science Foundation operated under cooperative agreement by Associated Universities, Inc.
 
\end{acknowledgements}

%
   \bibliographystyle{aa.bst}
   \bibliography{biblio}
%
%



\begin{appendix}

\clearpage

\section{The orientation of the jet axis}
\label{met_ori}

In \citet{Mos22} we have fixed the PA of the jet, PA$_{\rm jet}$, to be $\approx$50\degree\ on the basis of several evidences: \ 1)~the elongation at PA = 56\degree$\pm$12\degree\ of the slightly resolved (size $\approx$~150~au) JVLA continuum at 1.3~cm; \ 2)~the collimated, PA = 49\degree$\pm$18\degree, proper motions of the water masers near (100--200~au) the YSO from the previous VLBA (BeSSeL) observations; \ 3)~the direction  at PA $\approx$~44\degree\ of the non-thermal radio jet traced by the extended (size $\approx$~1000~au) JVLA continuum at 5~cm. On top of this evidence, we note that a change of \ $\gtrsim$5\degree\ in  PA$_{\rm jet}$ \ would compromise the sinusoidal fit of the maser streamlines. In fact, if the jet axis was oriented at 
 PA $\gtrsim$55\degree\ it would intersect the streamlines, while in correspondence of a jet PA $\lesssim$45\degree\ the lower and upper extremes of the streamlines would be placed at significant (and remarkably different) distances from the axis. Therefore, we can set \  PA$_{\rm jet} = 50$\degree$\pm5$\degree. 

Considering the small inclination with the plane of the sky of the maser 3D velocities from previous VLBA observations, in \citet{Mos22} we have set an upper limit for the jet inclination \ $i_{\rm jet} \le 30$\degree. Now, we improve on the estimate of the jet inclination by arguing that a selection of (5) masers from previous VLBA observations should accurately trace the jet direction in light of the following evidences: 1)~they are placed along the jet axis at the largest distance from the YSO; \ 2)~they move at relatively high speed, between 23 and 37~\kms; \ 3)~the PA of their proper motions is in the range 45\degree -- 55\degree;  \ 4)~the inclination angle of their 3D velocities varies in the small range \ 12\degree -- 19\degree, with an average inclination of 15\degree. Therefore, we can set \   $i_{\rm jet} = 15$\degree$\pm5$\degree, the jet axis being tilted toward us at PA = PA$_{\rm jet}$.

\section{Streamlines} 
\label{met_str}

\begin{figure}%
\includegraphics[width=0.49\textwidth]{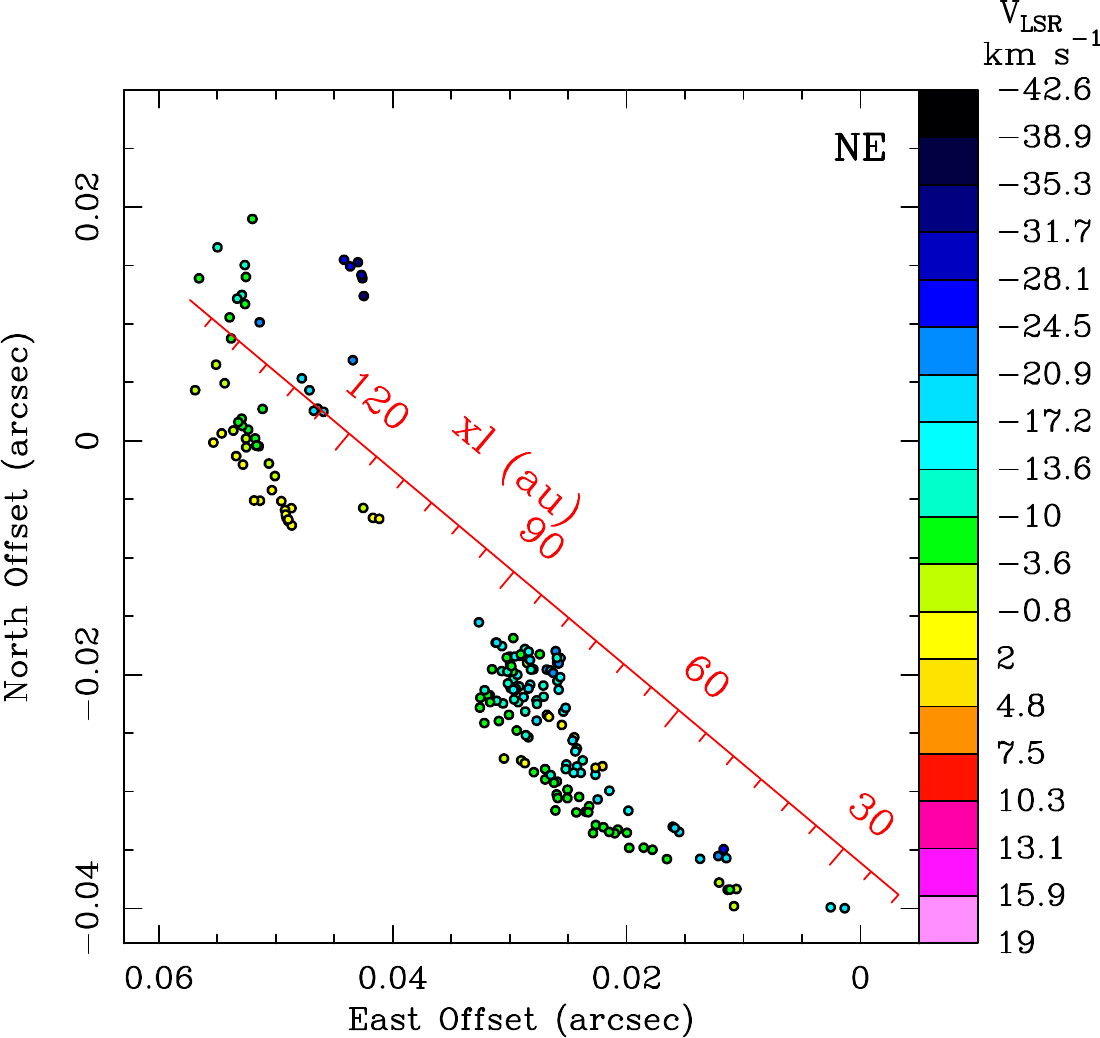}
\caption{{Non-simultaneous water masers.} Colored dots and the red axis have the same meaning as in Fig.~\ref{3V}. At variance with the left panel of Fig.~\ref{NE-SW}, all the detected masers are shown, including those observed only in a single epoch different from the first epoch. For these masers, missing the proper motion information, the first-epoch position cannot be derived.}
\label{ns-NE}
\end{figure}

In Figs.~\ref{3V}~and~\ref{NE-SW}, the plotted maser positions correspond to the first observing epoch, since we plot either masers detected at the first epoch or those observed at least in two successive epochs for which the first epoch position can be recovered through their proper motion. 
Instead, Fig.~\ref{ns-NE} shows all the masers detected across the four epochs.
The comparison between Fig.~\ref{ns-NE} and the left panel of Fig.~\ref{NE-SW} shows the importance of using simultaneous masers when tracing the DW streamlines. The positions of the masers observed only at either the second, third or fourth epoch fall in between the B1 and B2 streams, thus making it hard to recognize the two streamlines. 

Figure~\ref{Fover} shows that, in both the NE and SW regions, the older (2020) and newer (2023) maser patterns are spatially contiguous, intersect and partly overlap. As the maser streamlines change on small timescales \ $\lesssim$1~yr comparable or lower than the maser life time, the spatial and time overlap of different streamlines can make it difficult to discern the masers belonging to a given streamline.
In the following, we briefly describe the procedure employed to fit the three sinusoids to the maser positions tracing the R1, B1 and B2 streamlines. Our main criterion to select the masers is the quality of the sinusoidal fit. Since the parameters to fit are three, a streamline can be identified only if it is adequately sampled by a minimum of four maser clusters.

\begin{figure*}%
\includegraphics[width=0.52\textwidth]{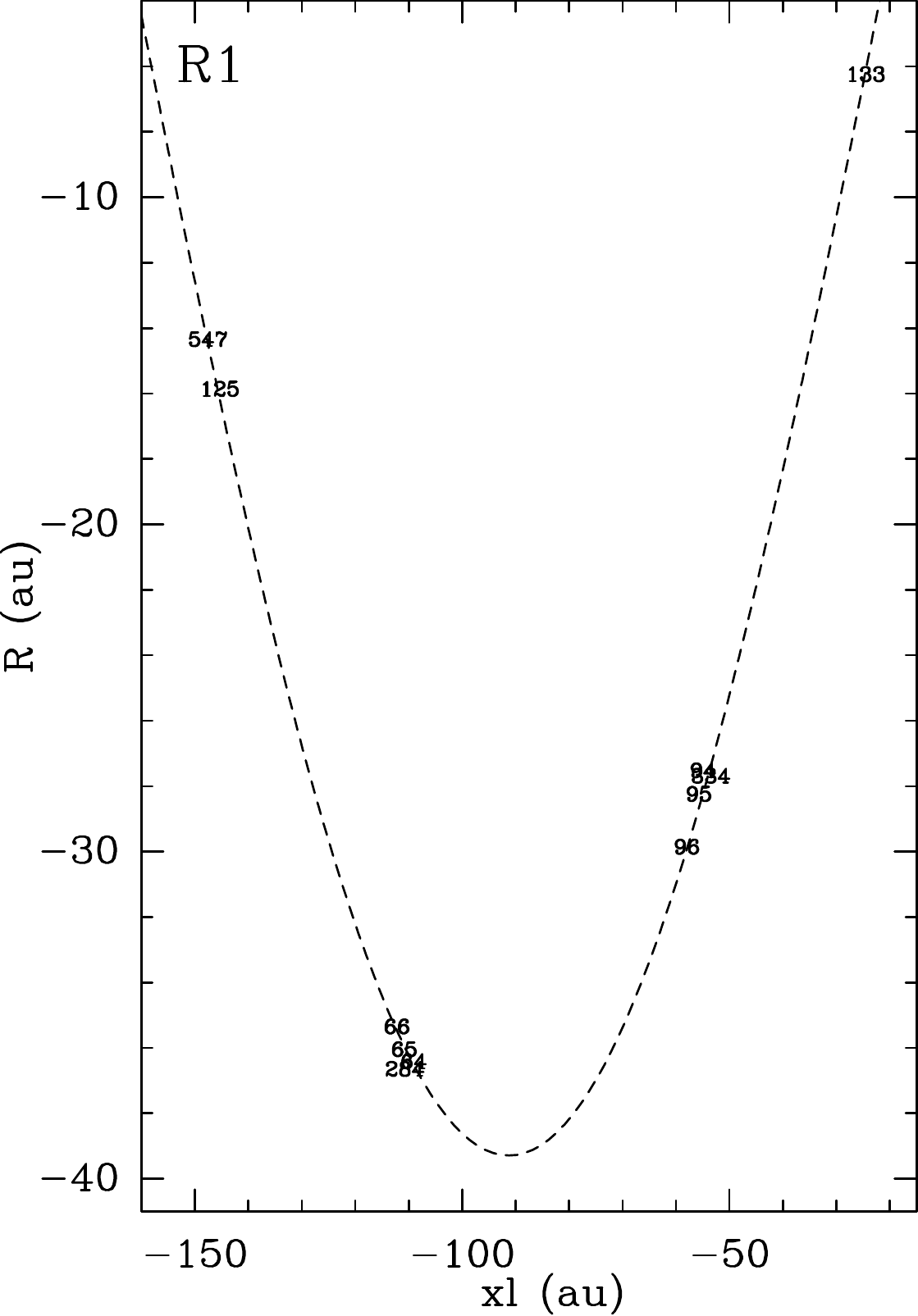}
\hspace*{0.5cm}\includegraphics[width=0.475\textwidth]{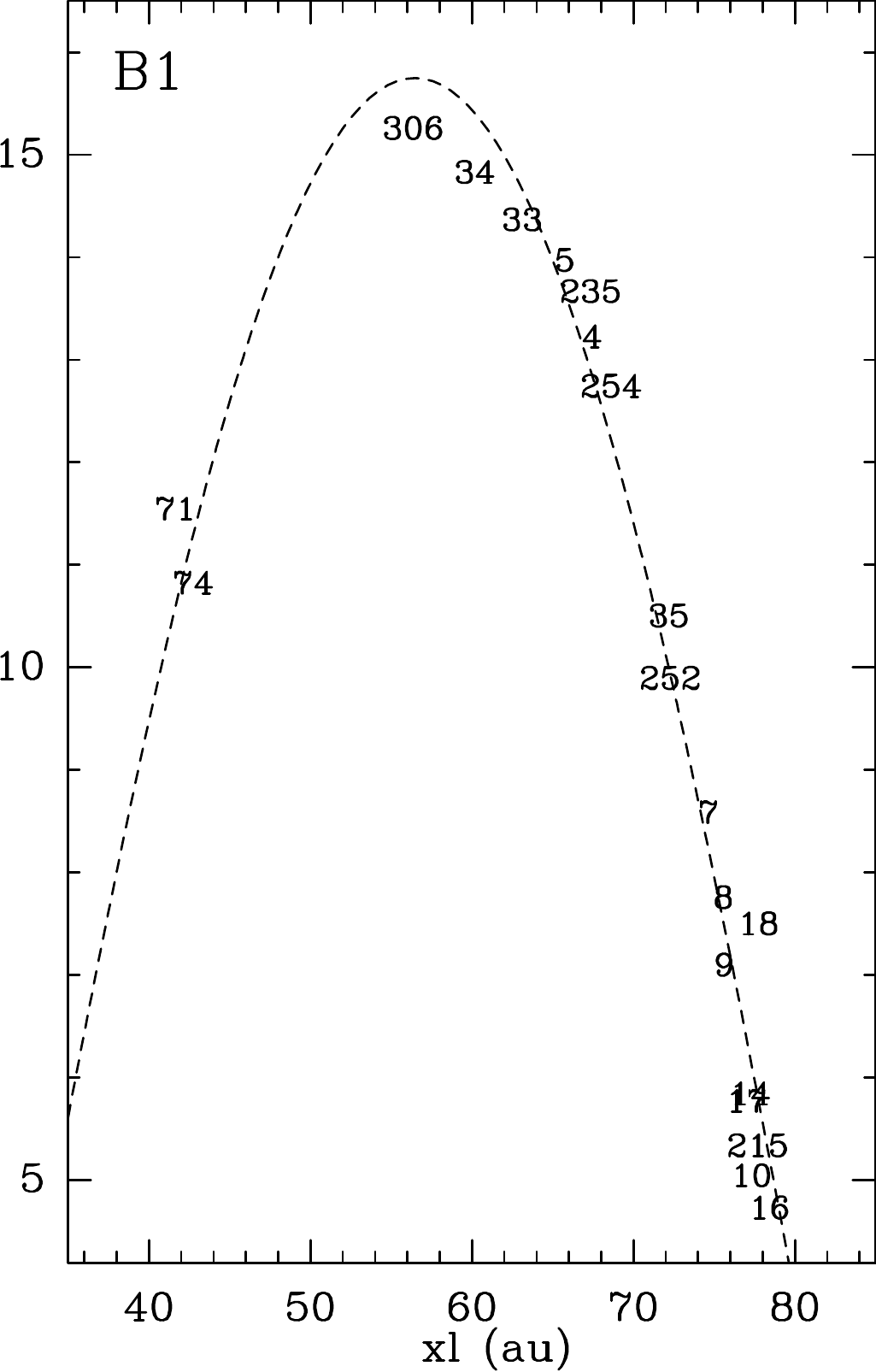}
\caption{{Sinusoidal fits of the R1 and B1 streamlines.} Maser labels (see Table~\ref{wat}) mark the position of the masers in the R1 (left~panel) and B1 (right) streamlines. We warn that the radial positions are stretched out to allow one to read the maser labels. Positional errors are not shown since they are smaller than the labels. In both panels, the black dashed line draws the sinusoid (see Table~\ref{tab_fit}) fitted to the maser positions.}
\label{R1B1}
\end{figure*}

As shown in the left panel of Fig.~\ref{R1B1}, the R1 streamline is traced by four, spatially distinct maser clusters. The sinusoidal fit of the maser positions has a very low rms deviation of \ 0.35~au. Besides, the sinusoidal fit of the positions of the three maser clusters closer to the YSO (over \ $-20$~au $\ge xl \ge$~$-120$~au) intercepts the fourth, more distant cluster (at $xl \approx -150$~au), too. Any other combination of masers of the SW region results in a sinusoidal fit of much lower quality. For instance, replacing the first two clusters of the R1 streamline (over \ $-20$~au~$\ge xl \ge$~$-60$~au) with those nearby but at slightly smaller radius (see Fig.~\ref{NE-SW}, right~panel), degrades the rms deviation of the sinusoidal fit to 1~au.

In the case of the B1 stream (see Fig.~\ref{R1B1}, right~panel), after excluding four masers clearly separated in \Vlsr\ (the yellow-green dots and arrows in Fig.~\ref{NE-SW}, left~panel), the remaining masers accurately sample (at more than 10 distinct positions) the sinusoid, whose fit has a low rms deviation of \ 0.65~au. Instead, in the case of the B2 stream, by including all the masers along the stream, the resulting sinusoidal fit has a significantly higher rms of \ 1.4~au. Differently from the R1 stream, the number of masers in the B2 stream is too high to select unambiguously the positions that provide the best sinusoidal fit. Therefore, we have considered the velocities to see whether it was possible to distinguish masers belonging to different streamlines on the basis of their motion.

\begin{figure}%
\includegraphics[width=0.49\textwidth]{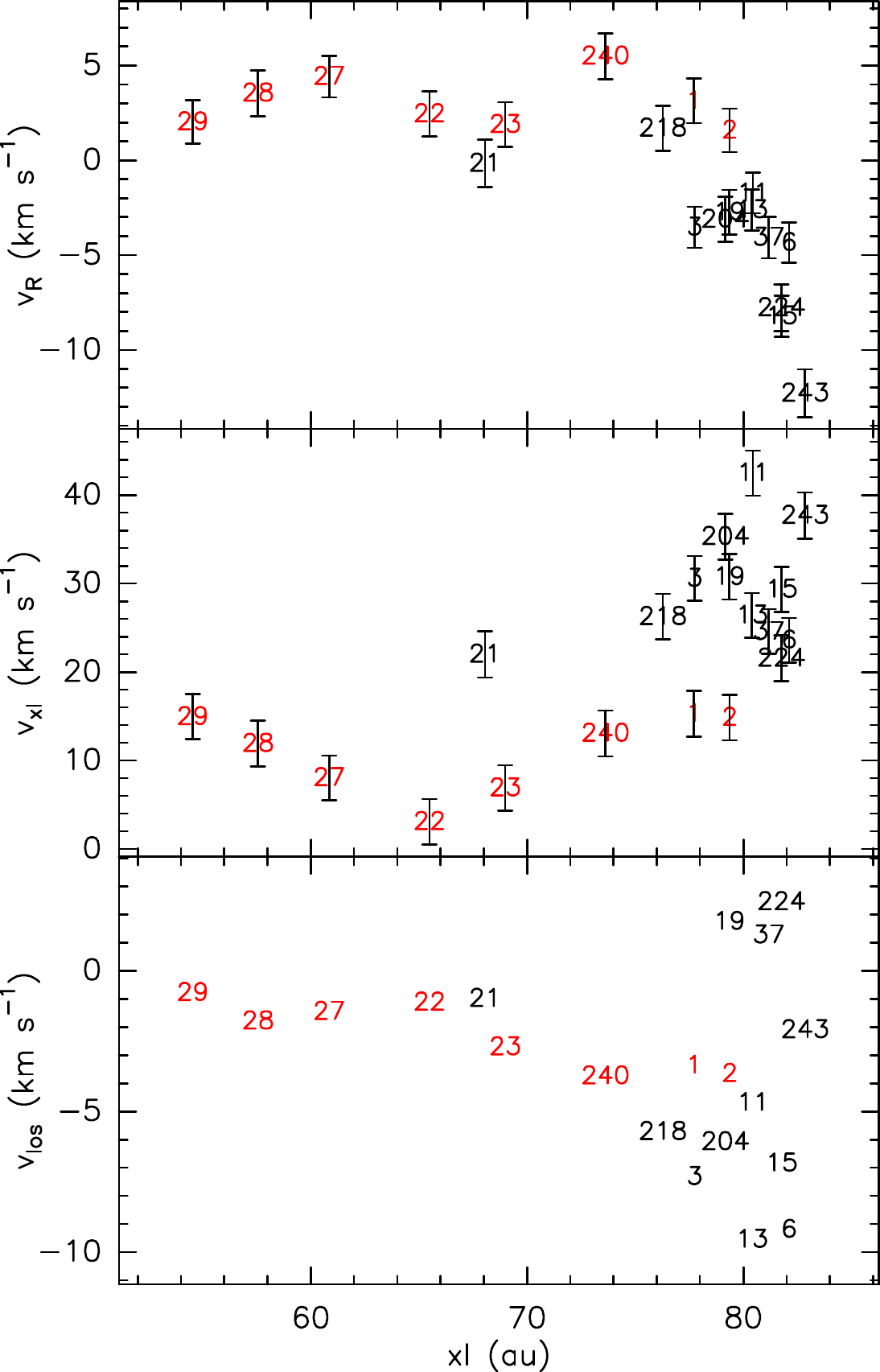}
\caption{{Distribution of velocities in the B2 stream.} Maser labels (see Table~\ref{wat}) and error bars report values and corresponding errors for the velocity components along the \ $R$ (upper panel), $xl$ (middle panel) and \ los (lower panel) axes. The error bars of the los components are not shown, as they are smaller than the labels. Red labels denote a subset of masers with similar velocity components.}
\label{B2v}
\end{figure}

\begin{figure}%
\includegraphics[width=0.49\textwidth]{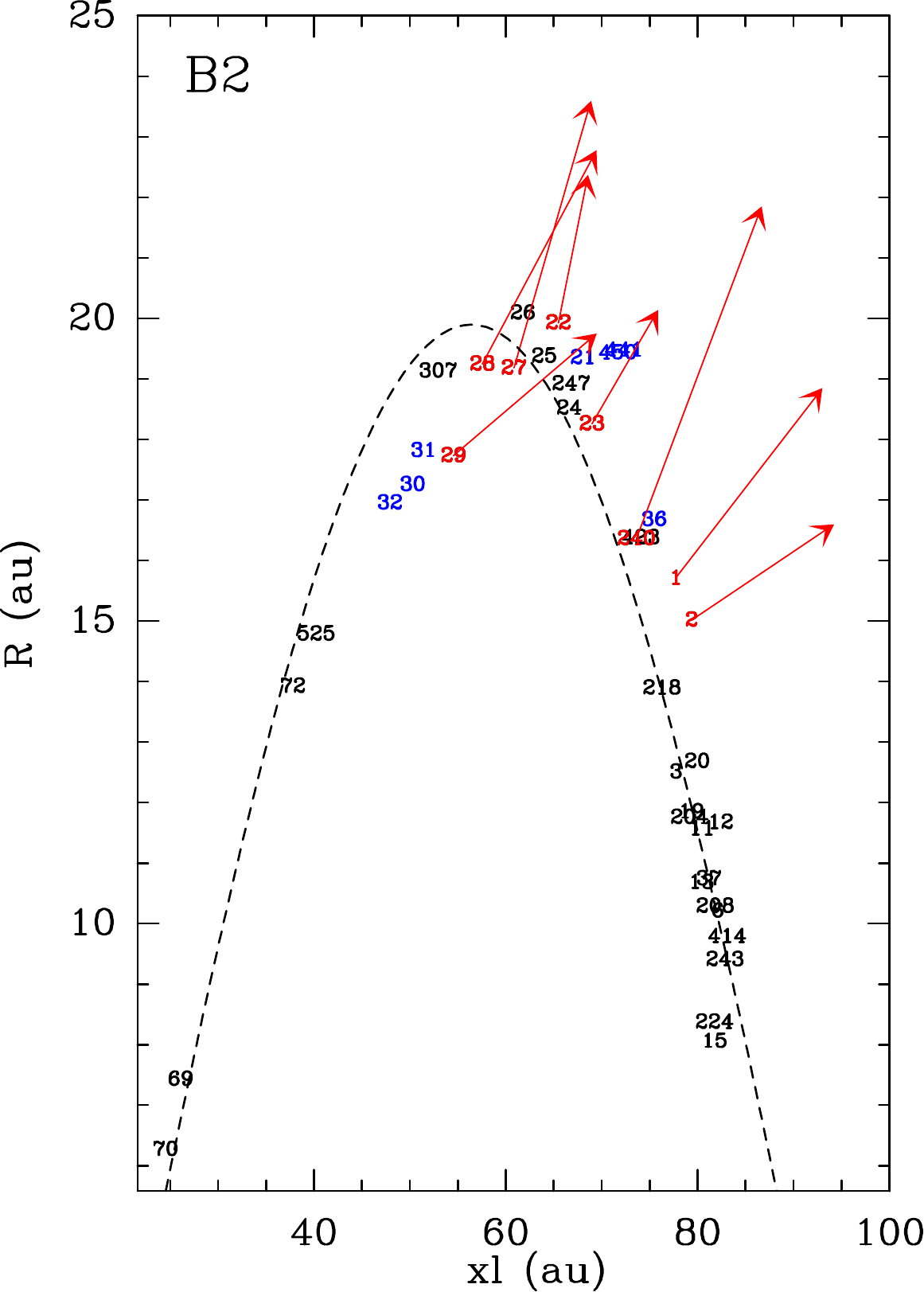}
\caption{{Sinusoidal fit of the B2 streamline.} Maser labels (see Table~\ref{wat}) mark the position of the masers of the B2 stream. We warn that the radial positions are stretched out to allow one to read the maser labels. Positional errors are not shown since they are smaller than the labels. Red labels and arrows denote the positions and proper motions of a subset of masers showing similar velocities (see Fig.~\ref{B2v}). The black dashed line draws the sinusoid (see Table~\ref{tab_fit}) fitted to the positions of the masers with black labels. The blue labels indicate a few masers scattered in position with respect to the sinusoid.}
\label{B2_sel}
\end{figure}

Figure~\ref{B2v} shows that a set of (8) masers (red labels) of the B2 stream have almost the same velocities in comparison with the rest of the masers presenting a much higher velocity scatter. Figure~\ref{B2_sel} shows the proper motions (red arrows) of these eight masers, which are about parallel to each other. Assuming, conservatively, that this set of masers belongs to a different streamline, we have excluded them and repeated the sinusoidal fit of the B2 streamline. The result is presented in Fig.~\ref{B2_sel}, which shows that a few (blue labels) other masers can be also excluded on the basis of the very scattered positions. The sinusoidal fit to the remaining masers (black labels), defining the B2 streamline, has a rms deviation of 0.9~au.

Finally, we have also tried to fit a sinusoid to the masers of the B3 stream inside the NE region (see Fig.\ref{NE-SW}, left~panel) whose spatial distribution is roughly arc-like. After excluding a few, most scattered masers, the rms deviation of the sinusoidal fit is 2~au, 2--6 times larger than that of the B2, B1 and R1 streamlines. Besides, the analysis of the maser velocities does not provide any clue to separate the masers into subsets belonging to different streamlines. Our conclusion is that we cannot employ these masers to identify a streamline with a confidence level as good as that of the previous three streamlines.  

\FloatBarrier

\section{The reference system co-moving with the young stellar object}
\label{met_ref}

As described in Sect.~\ref{met_obs}, the relative proper motions are significantly more accurate than the absolute proper motions. However, to use them to describe the motion with respect to the YSO, we need to determine the proper motion of the ``center of motion" (feature \#0) with respect to the YSO. A good reference for that is the absolute proper motion of feature \#0: $v_{\rm RA} = 4.5\pm1.7$~\kms, $v_{\rm DEC} = 18.5\pm1.7$~\kms\ (or, equivalently, $v_{xl} = 15.3\pm1.7$~\kms, $v_{R} = -11.3\pm1.7$~\kms). Although this proper motion is corrected for the parallax, solar motion and differential Galactic rotation, the virial motion of an individual YSO can still deviate from that of the harboring high-mass star-forming region up to $\approx$10~\kms\ \citep{Rei19}. In the following, we derive an independent estimate of the proper motion of feature \#0 with respect to the YSO accurate within a few kilometers per second.

While the relative proper motions are calculated with respect to feature \#0, the good knowledge of the YSO's \Vlsr\ allows us to determine accurate values of the velocity component along the line-of-sight, $v_{\rm los}$. If for a set of masers \ $v_{\rm los}$ \ can be expected to be directly proportional to one of the two proper motion components, $v_{\rm RA}$ or  $v_{\rm DEC}$ (or, equivalently, $v_{xl}$ or  $v_{R}$), the offset of that component can be determined.

First, we apply this method to the set of eight masers of the B2 stream whose proper motions are almost parallel (see Fig.~\ref{B2_sel}). The 3D velocities of these masers are also approximately parallel, as we have verified that they are  inclined on average by only \ 13\degree\ with respect to a central direction. Let us indicate with \ $M$ and $\delta$ \ the mean and standard deviation, respectively, of the values of one specific velocity component. We can expect that: $M_{xl} / M_{\rm los} = \delta_{xl} / \delta_{\rm los} $ and $M_{R} / M_{\rm los} = \delta_{R} / \delta_{\rm los} $. Since the standard deviations are not affected by the presence of an offset, these relations allow us to calculate the corrected value of  \ $M_{xl}$ \ and $M_{R}$, and evaluate the proper motion of feature \#0 with respect to the YSO. Using all or a subset of the eight masers, we always find consistent corrections resulting into velocity components of feature \#0 varying within the small ranges:  $v_{xl}^{f0} = 17.5\pm2.5$~\kms\ and \ $v_{R}^{f0} = -12.5\pm1.0$~\kms. We note that the derived correction agrees with the absolute proper motion of feature \#0 reported above. 

As discussed in Sect.~\ref{Dis_MP}, the masers belonging to the N region move radially away at a similar azimuthal angle of $\approx$304\degree. Since \ $v_{\rho}$ \ dominates on the other velocity components (see Table~\ref{tab_vel}), $v_{\rm los}$ and $v_{R}$ \ are expected to be directly proportional. Using the relation: $M_{R} / M_{\rm los} = \delta_{R} / \delta_{\rm los} $, we derive  \ $v_{R}^{f0} = -11.5$~\kms, consistent with the previous determination.
In conclusion, the good agreement among these different methods make us confident that the corrected relative proper motions represent the velocities in the reference frame of the YSO with accuracies of a few kilometers per second.

\end{appendix}
    

\setcounter{table}{0}

\onecolumn
\begin{longtable}{ccccrrrr} 
\caption{\label{wat} 22.2~GHz H$_2$O maser parameters for \targ.}\\  
\hline\hline
Feature & Epochs\tablefootmark{a} of & I$_{\rm peak}$ & $V_{\rm LSR}$ & \multicolumn{1}{c}{$\Delta~x$} & \multicolumn{1}{c}{$\Delta~y$} & \multicolumn{1}{c}{$V_{x}$} & \multicolumn{1}{c}{$V_{y}$} \\
Number  & Detection & (Jy beam$^{-1}$) & (km s$^{-1}$) & \multicolumn{1}{c}{(mas)} & \multicolumn{1}{c}{(mas)} & \multicolumn{1}{c}{(km s$^{-1}$)} & \multicolumn{1}{c}{(km s$^{-1}$)} \\
\hline
\endfirsthead
\caption{continued.}\\
\hline\hline
Feature & Epochs\tablefootmark{a} of & I$_{\rm peak}$ & $V_{\rm LSR}$ & \multicolumn{1}{c}{$\Delta~x$} & \multicolumn{1}{c}{$\Delta~y$} & \multicolumn{1}{c}{$V_{x}$} & \multicolumn{1}{c}{$V_{y}$} \\
Number  & Detection & (Jy beam$^{-1}$) & (km s$^{-1}$) & \multicolumn{1}{c}{(mas)} & \multicolumn{1}{c}{(mas)} & \multicolumn{1}{c}{(km s$^{-1}$)} & \multicolumn{1}{c}{(km s$^{-1}$)} \\
\hline
\endhead
\hline
\endfoot
\hline
\endlastfoot
      0 &        1,2,3,4 &    83.559 &   -17.8 &     0.00\hspace*{0.4cm} &     0.00\hspace*{0.4cm} &    0.0\hspace*{0.5cm} &    0.0\hspace*{0.5cm} \\
    204 &          2,3,4 &    76.447 &   -12.5 &    29.82$\pm$0.02 &   -20.92$\pm$0.02 &   19.7$\pm$ 0.6 &    4.2$\pm$ 0.6  \\
      6 &        1,2,3,4 &    71.077 &   -15.6 &    29.96$\pm$0.02 &   -19.12$\pm$0.02 &    9.9$\pm$ 0.4 &   -2.3$\pm$ 0.4  \\
    215 &            2,3 &    52.493 &   -13.0 &    27.05$\pm$0.02 &   -18.69$\pm$0.02 &    ....\hspace*{0.5cm} &    ....\hspace*{0.5cm}  \\
    205 &              2 &    48.506 &   -13.9 &    29.45$\pm$0.02 &   -21.28$\pm$0.02 &    ....\hspace*{0.5cm} &    ....\hspace*{0.5cm}  \\
      7 &        1,2,3,4 &    47.059 &   -14.0 &    25.81$\pm$0.02 &   -21.29$\pm$0.02 &   32.4$\pm$ 0.4 &   -2.1$\pm$ 0.4  \\
      4 &        1,2,3,4 &    33.967 &   -12.3 &    24.25$\pm$0.02 &   -26.32$\pm$0.02 &   31.3$\pm$ 0.4 &   -6.3$\pm$ 0.4  \\
      1 &          1,2,3 &    27.985 &    -9.8 &    30.06$\pm$0.02 &   -23.44$\pm$0.02 &    8.4$\pm$ 0.6 &  -13.4$\pm$ 0.6  \\
    613 &              4 &    22.315 &   -15.9 &    28.24$\pm$0.02 &   -20.86$\pm$0.02 &    ....\hspace*{0.5cm} &    ....\hspace*{0.5cm}  \\
    280 &          2,3,4 &    17.278 &   -27.7 &    44.19$\pm$0.02 &    15.49$\pm$0.02 &   17.3$\pm$ 0.7 &   17.2$\pm$ 0.7  \\
      3 &        1,2,3,4 &    17.125 &   -13.7 &    28.81$\pm$0.02 &   -21.92$\pm$0.02 &   15.8$\pm$ 0.4 &    1.5$\pm$ 0.4  \\
    257 &            2,3 &    15.481 &    -3.8 &    51.72$\pm$0.02 &    -0.23$\pm$0.02 &    ....\hspace*{0.5cm} &    ....\hspace*{0.5cm}  \\
    207 &              2 &    14.196 &   -12.2 &    30.19$\pm$0.02 &   -19.74$\pm$0.02 &    ....\hspace*{0.5cm} &    ....\hspace*{0.5cm}  \\
      9 &              1 &    14.055 &   -15.0 &    25.67$\pm$0.02 &   -20.22$\pm$0.02 &    ....\hspace*{0.5cm} &    ....\hspace*{0.5cm}  \\
    202 &              2 &    13.515 &   -10.8 &    27.09$\pm$0.02 &   -21.88$\pm$0.02 &    ....\hspace*{0.5cm} &    ....\hspace*{0.5cm}  \\
    643 &              4 &    12.919 &   -31.1 &    61.87$\pm$0.02 &    59.08$\pm$0.02 &    ....\hspace*{0.5cm} &    ....\hspace*{0.5cm}  \\
    217 &              2 &    12.690 &   -12.4 &    29.27$\pm$0.02 &   -22.36$\pm$0.02 &    ....\hspace*{0.5cm} &    ....\hspace*{0.5cm}  \\
    208 &            2,3 &    11.659 &    -6.6 &    30.31$\pm$0.02 &   -19.15$\pm$0.02 &    ....\hspace*{0.5cm} &    ....\hspace*{0.5cm}  \\
    423 &            3,4 &    11.354 &   -10.9 &    29.35$\pm$0.02 &   -25.20$\pm$0.02 &    ....\hspace*{0.5cm} &    ....\hspace*{0.5cm}  \\
    264 &              2 &    10.810 &     0.1 &    49.21$\pm$0.02 &    -5.91$\pm$0.02 &    ....\hspace*{0.5cm} &    ....\hspace*{0.5cm}  \\
     13 &        1,2,3,4 &    10.784 &   -15.9 &    29.34$\pm$0.02 &   -20.02$\pm$0.02 &   13.2$\pm$ 0.4 &   -1.9$\pm$ 0.4  \\
     69 &              1 &    10.262 &   -15.9 &     2.54$\pm$0.03 &   -39.90$\pm$0.03 &    ....\hspace*{0.5cm} &    ....\hspace*{0.5cm}  \\
     14 &              1 &     8.940 &   -11.3 &    25.94$\pm$0.02 &   -18.98$\pm$0.02 &    ....\hspace*{0.5cm} &    ....\hspace*{0.5cm}  \\
     41 &        1,2,3,4 &     8.731 &   -31.5 &    42.59$\pm$0.02 &    13.91$\pm$0.02 &   13.6$\pm$ 0.4 &   14.6$\pm$ 0.4  \\
    618 &              4 &     8.608 &    -9.9 &    30.92$\pm$0.02 &   -23.97$\pm$0.02 &    ....\hspace*{0.5cm} &    ....\hspace*{0.5cm}  \\
     55 &        1,2,3,4 &     8.432 &    -0.0 &    49.53$\pm$0.03 &    -5.16$\pm$0.03 &    1.0$\pm$ 0.4 &   -6.3$\pm$ 0.4  \\
     15 &        1,2,3,4 &     8.193 &   -13.2 &    28.95$\pm$0.02 &   -18.24$\pm$0.02 &   11.8$\pm$ 0.4 &    4.3$\pm$ 0.4  \\
    224 &          2,3,4 &     8.185 &    -4.0 &    29.24$\pm$0.02 &   -18.44$\pm$0.02 &    6.1$\pm$ 0.7 &   -1.0$\pm$ 0.7  \\
     58 &              1 &     8.068 &    -3.6 &    51.46$\pm$0.02 &    -0.47$\pm$0.02 &    ....\hspace*{0.5cm} &    ....\hspace*{0.5cm}  \\
    414 &            3,4 &     7.879 &    -6.8 &    30.69$\pm$0.02 &   -18.51$\pm$0.02 &    ....\hspace*{0.5cm} &    ....\hspace*{0.5cm}  \\
    283 &              2 &     7.571 &    -4.7 &   -79.28$\pm$0.02 &   -73.35$\pm$0.02 &    ....\hspace*{0.5cm} &    ....\hspace*{0.5cm}  \\
     11 &        1,2,3,4 &     5.979 &   -11.1 &    29.71$\pm$0.02 &   -20.41$\pm$0.02 &   26.1$\pm$ 0.4 &    7.8$\pm$ 0.4  \\
     64 &          1,2,3 &     5.855 &    -4.7 &   -78.28$\pm$0.02 &   -72.59$\pm$0.02 &  -19.0$\pm$ 0.7 &   -8.3$\pm$ 0.7  \\
      2 &          1,2,3 &     5.640 &   -10.1 &    30.57$\pm$0.02 &   -22.46$\pm$0.02 &    7.0$\pm$ 0.6 &  -12.5$\pm$ 0.6  \\
    607 &              4 &     5.255 &   -10.1 &    32.14$\pm$0.02 &   -21.33$\pm$0.02 &    ....\hspace*{0.5cm} &    ....\hspace*{0.5cm}  \\
    218 &          2,3,4 &     4.948 &   -12.1 &    29.16$\pm$0.02 &   -23.34$\pm$0.02 &   15.8$\pm$ 0.7 &   -5.2$\pm$ 0.7  \\
     45 &        1,2,3,4 &     4.944 &   -12.0 &    52.91$\pm$0.03 &    12.48$\pm$0.03 &   11.2$\pm$ 0.5 &    5.0$\pm$ 0.5  \\
     57 &          1,2,3 &     4.513 &    -0.2 &    50.34$\pm$0.02 &    -4.22$\pm$0.02 &   -0.7$\pm$ 0.6 &   -8.4$\pm$ 0.6  \\
    223 &              2 &     4.030 &   -15.1 &    26.51$\pm$0.02 &   -19.65$\pm$0.02 &    ....\hspace*{0.5cm} &    ....\hspace*{0.5cm}  \\
    254 &          2,3,4 &     3.494 &   -17.6 &    25.65$\pm$0.03 &   -25.51$\pm$0.04 &   32.5$\pm$ 0.8 &    5.0$\pm$ 0.9  \\
      8 &        1,2,3,4 &     3.359 &   -13.2 &    25.91$\pm$0.02 &   -20.54$\pm$0.02 &   26.3$\pm$ 0.4 &   19.3$\pm$ 0.4  \\
    235 &          2,3,4 &     3.201 &   -15.2 &    25.04$\pm$0.02 &   -26.92$\pm$0.02 &   19.8$\pm$ 0.6 &  -12.2$\pm$ 0.6  \\
    490 &              3 &     3.085 &   -33.2 &   -16.47$\pm$0.02 &   -25.20$\pm$0.02 &    ....\hspace*{0.5cm} &    ....\hspace*{0.5cm}  \\
    260 &          2,3,4 &     2.972 &    -3.0 &    52.20$\pm$0.02 &    -0.69$\pm$0.03 &   -9.3$\pm$ 0.7 &   -4.8$\pm$ 0.7  \\
    427 &              3 &     2.786 &   -15.3 &    25.15$\pm$0.02 &   -27.68$\pm$0.02 &    ....\hspace*{0.5cm} &    ....\hspace*{0.5cm}  \\
    413 &              3 &     2.616 &    -9.3 &    31.19$\pm$0.02 &   -17.26$\pm$0.02 &    ....\hspace*{0.5cm} &    ....\hspace*{0.5cm}  \\
     12 &              1 &     2.576 &   -10.6 &    30.68$\pm$0.02 &   -19.69$\pm$0.02 &    ....\hspace*{0.5cm} &    ....\hspace*{0.5cm}  \\
    421 &              3 &     2.568 &   -15.1 &    29.68$\pm$0.02 &   -21.29$\pm$0.02 &    ....\hspace*{0.5cm} &    ....\hspace*{0.5cm}  \\
     73 &        1,2,3,4 &     2.539 &   -27.4 &   -37.02$\pm$0.02 &    37.19$\pm$0.02 &  -26.2$\pm$ 0.4 &    7.3$\pm$ 0.4  \\
     65 &          1,3,4 &     2.497 &    -4.8 &   -78.96$\pm$0.03 &   -73.45$\pm$0.02 &  -20.8$\pm$ 0.4 &   -9.1$\pm$ 0.4  \\
     16 &        1,2,3,4 &     2.237 &   -22.8 &    26.06$\pm$0.02 &   -17.99$\pm$0.02 &   27.7$\pm$ 0.4 &    4.4$\pm$ 0.4  \\
      5 &          1,2,3 &     2.156 &   -11.7 &    23.75$\pm$0.02 &   -27.33$\pm$0.02 &   22.5$\pm$ 0.6 &   -6.6$\pm$ 0.6  \\
    510 &            3,4 &     2.043 &   -28.2 &   -44.20$\pm$0.02 &    36.64$\pm$0.02 &    ....\hspace*{0.5cm} &    ....\hspace*{0.5cm}  \\
    461 &              3 &     1.910 &     0.4 &    49.15$\pm$0.02 &    -6.30$\pm$0.02 &    ....\hspace*{0.5cm} &    ....\hspace*{0.5cm}  \\
     42 &          1,2,3 &     1.833 &   -32.4 &    42.47$\pm$0.02 &    12.39$\pm$0.02 &   16.5$\pm$ 0.6 &   14.8$\pm$ 0.6  \\
    231 &              2 &     1.743 &     0.6 &    26.64$\pm$0.02 &   -23.62$\pm$0.02 &    ....\hspace*{0.5cm} &    ....\hspace*{0.5cm}  \\
     84 &            1,2 &     1.726 &   -31.7 &    59.83$\pm$0.02 &    55.22$\pm$0.02 &    ....\hspace*{0.5cm} &    ....\hspace*{0.5cm}  \\
     39 &          1,2,3 &     1.711 &     1.5 &    25.54$\pm$0.03 &   -24.30$\pm$0.03 &   19.9$\pm$ 0.7 &    8.4$\pm$ 0.7  \\
    629 &              4 &     1.700 &   -14.8 &    28.37$\pm$0.02 &   -18.03$\pm$0.02 &    ....\hspace*{0.5cm} &    ....\hspace*{0.5cm}  \\
    290 &              2 &     1.691 &    11.3 &   -36.88$\pm$0.02 &   -62.80$\pm$0.02 &    ....\hspace*{0.5cm} &    ....\hspace*{0.5cm}  \\
    465 &            3,4 &     1.677 &   -18.9 &    47.25$\pm$0.02 &     3.32$\pm$0.02 &    ....\hspace*{0.5cm} &    ....\hspace*{0.5cm}  \\
     17 &          1,2,3 &     1.624 &   -23.1 &    25.80$\pm$0.02 &   -19.04$\pm$0.02 &   31.2$\pm$ 0.6 &   17.7$\pm$ 0.6  \\
    105 &        1,2,3,4 &     1.615 &   -15.2 &   -52.19$\pm$0.04 &    19.84$\pm$0.04 &  -19.5$\pm$ 0.5 &   -4.7$\pm$ 0.5  \\
    632 &              4 &     1.531 &    -9.1 &    29.05$\pm$0.03 &   -18.26$\pm$0.03 &    ....\hspace*{0.5cm} &    ....\hspace*{0.5cm}  \\
     37 &        1,2,3,4 &     1.489 &    -5.1 &    29.73$\pm$0.03 &   -19.74$\pm$0.03 &   10.8$\pm$ 0.4 &   -1.9$\pm$ 0.4  \\
    250 &              2 &     1.456 &   -14.5 &    29.55$\pm$0.02 &   -18.45$\pm$0.02 &    ....\hspace*{0.5cm} &    ....\hspace*{0.5cm}  \\
    236 &              2 &     1.439 &   -11.2 &    28.70$\pm$0.02 &   -17.80$\pm$0.02 &    ....\hspace*{0.5cm} &    ....\hspace*{0.5cm}  \\
    303 &              2 &     1.414 &    -6.8 &    20.74$\pm$0.02 &   -33.25$\pm$0.02 &    ....\hspace*{0.5cm} &    ....\hspace*{0.5cm}  \\
    274 &          2,3,4 &     1.408 &    -0.8 &    52.51$\pm$0.02 &    -2.59$\pm$0.03 &   -9.1$\pm$ 0.7 &  -17.3$\pm$ 0.7  \\
    480 &            3,4 &     1.361 &   -30.4 &    43.57$\pm$0.02 &    15.27$\pm$0.02 &    ....\hspace*{0.5cm} &    ....\hspace*{0.5cm}  \\
    229 &              2 &     1.228 &   -14.6 &    24.25$\pm$0.02 &   -27.86$\pm$0.02 &    ....\hspace*{0.5cm} &    ....\hspace*{0.5cm}  \\
    240 &          2,3,4 &     1.220 &   -10.1 &    28.65$\pm$0.02 &   -25.86$\pm$0.02 &    8.2$\pm$ 0.7 &  -16.6$\pm$ 0.7  \\
     59 &          1,2,3 &     1.215 &    -3.0 &    52.55$\pm$0.02 &     0.20$\pm$0.02 &   -1.5$\pm$ 0.6 &   -7.5$\pm$ 0.6  \\
     19 &          1,2,3 &     1.204 &    -4.7 &    29.31$\pm$0.02 &   -20.98$\pm$0.02 &   16.4$\pm$ 0.6 &    1.0$\pm$ 0.6  \\
    633 &              4 &     1.194 &   -17.7 &    31.12$\pm$0.02 &   -17.26$\pm$0.02 &    ....\hspace*{0.5cm} &    ....\hspace*{0.5cm}  \\
     23 &          1,2,3 &     1.157 &    -9.1 &    26.97$\pm$0.02 &   -28.08$\pm$0.02 &    1.1$\pm$ 0.6 &  -17.8$\pm$ 0.6  \\
     93 &              1 &     1.153 &     2.4 &   -50.07$\pm$0.02 &   -57.33$\pm$0.02 &    ....\hspace*{0.5cm} &    ....\hspace*{0.5cm}  \\
    203 &              2 &     1.095 &   -11.2 &    27.12$\pm$0.02 &   -20.93$\pm$0.02 &    ....\hspace*{0.5cm} &    ....\hspace*{0.5cm}  \\
    113 &              1 &     1.091 &   -21.6 &   -15.03$\pm$0.02 &   -37.29$\pm$0.02 &    ....\hspace*{0.5cm} &    ....\hspace*{0.5cm}  \\
    713 &              4 &     1.076 &   -29.2 &   -45.85$\pm$0.02 &     4.57$\pm$0.02 &    ....\hspace*{0.5cm} &    ....\hspace*{0.5cm}  \\
     10 &              1 &     1.057 &   -15.1 &    25.66$\pm$0.03 &   -18.58$\pm$0.02 &    ....\hspace*{0.5cm} &    ....\hspace*{0.5cm}  \\
    430 &              3 &     1.054 &    -9.4 &    31.88$\pm$0.02 &   -21.75$\pm$0.02 &    ....\hspace*{0.5cm} &    ....\hspace*{0.5cm}  \\
     52 &        1,2,3,4 &     1.052 &   -18.3 &    46.43$\pm$0.07 &     2.76$\pm$0.05 &    2.4$\pm$ 0.6 &   -2.2$\pm$ 0.6  \\
    716 &              4 &     1.008 &    -3.4 &    10.81$\pm$0.02 &   -39.81$\pm$0.02 &    ....\hspace*{0.5cm} &    ....\hspace*{0.5cm}  \\
     86 &        1,2,3,4 &     1.004 &   -32.2 &    60.63$\pm$0.02 &    57.72$\pm$0.02 &   18.8$\pm$ 0.4 &   31.5$\pm$ 0.4  \\
     89 &            1,2 &     1.004 &   -22.3 &   -27.16$\pm$0.02 &   -23.29$\pm$0.02 &    ....\hspace*{0.5cm} &    ....\hspace*{0.5cm}  \\
    324 &              2 &     0.997 &   -29.6 &   -15.91$\pm$0.02 &   -25.43$\pm$0.02 &    ....\hspace*{0.5cm} &    ....\hspace*{0.5cm}  \\
    441 &            3,4 &     0.994 &    -1.7 &    29.85$\pm$0.03 &   -27.72$\pm$0.04 &    ....\hspace*{0.5cm} &    ....\hspace*{0.5cm}  \\
     56 &              1 &     0.987 &    -0.2 &    48.67$\pm$0.02 &    -5.76$\pm$0.02 &    ....\hspace*{0.5cm} &    ....\hspace*{0.5cm}  \\
    258 &          2,3,4 &     0.984 &    -4.7 &    52.51$\pm$0.03 &     0.73$\pm$0.03 &    4.4$\pm$ 0.7 &   -6.9$\pm$ 0.8  \\
    230 &              2 &     0.976 &   -15.3 &    23.92$\pm$0.02 &   -28.40$\pm$0.02 &    ....\hspace*{0.5cm} &    ....\hspace*{0.5cm}  \\
    334 &          2,3,4 &     0.949 &    -2.1 &   -49.44$\pm$0.02 &   -55.61$\pm$0.03 &  -22.3$\pm$ 0.7 &  -26.1$\pm$ 0.7  \\
    114 &              1 &     0.945 &   -14.5 &   -51.90$\pm$0.02 &     8.31$\pm$0.02 &    ....\hspace*{0.5cm} &    ....\hspace*{0.5cm}  \\
     60 &        1,2,3,4 &     0.939 &    -2.7 &    50.59$\pm$0.02 &    -1.94$\pm$0.02 &    2.6$\pm$ 0.4 &   -2.1$\pm$ 0.4  \\
    431 &              3 &     0.934 &    -9.4 &    31.67$\pm$0.02 &   -22.35$\pm$0.02 &    ....\hspace*{0.5cm} &    ....\hspace*{0.5cm}  \\
     26 &            1,2 &     0.917 &    -7.6 &    24.30$\pm$0.02 &   -31.78$\pm$0.02 &    ....\hspace*{0.5cm} &    ....\hspace*{0.5cm}  \\
    243 &          2,3,4 &     0.915 &    -8.5 &    30.53$\pm$0.02 &   -18.01$\pm$0.02 &   15.6$\pm$ 0.8 &   12.8$\pm$ 0.8  \\
    519 &              3 &     0.893 &    -7.4 &    21.01$\pm$0.02 &   -33.57$\pm$0.02 &    ....\hspace*{0.5cm} &    ....\hspace*{0.5cm}  \\
    284 &          2,3,4 &     0.883 &    -5.1 &   -79.86$\pm$0.02 &   -73.89$\pm$0.02 &  -22.6$\pm$ 1.0 &  -24.2$\pm$ 1.0  \\
    483 &            3,4 &     0.865 &    -5.6 &    56.77$\pm$0.02 &    14.55$\pm$0.02 &    ....\hspace*{0.5cm} &    ....\hspace*{0.5cm}  \\
     44 &          1,2,3 &     0.863 &   -11.6 &    52.65$\pm$0.02 &    15.04$\pm$0.02 &   15.2$\pm$ 0.6 &   10.4$\pm$ 0.6  \\
     28 &          1,3,4 &     0.850 &    -8.2 &    21.98$\pm$0.06 &   -33.05$\pm$0.05 &    6.0$\pm$ 0.7 &  -15.9$\pm$ 0.6  \\
    428 &              3 &     0.848 &   -15.5 &    24.52$\pm$0.02 &   -28.39$\pm$0.02 &    ....\hspace*{0.5cm} &    ....\hspace*{0.5cm}  \\
     34 &            1,2 &     0.814 &   -16.9 &    21.47$\pm$0.03 &   -29.93$\pm$0.03 &    ....\hspace*{0.5cm} &    ....\hspace*{0.5cm}  \\
     91 &          1,2,3 &     0.811 &     3.2 &   -50.36$\pm$0.03 &   -58.31$\pm$0.02 &  -30.9$\pm$ 0.6 &  -26.3$\pm$ 0.6  \\
     98 &            1,2 &     0.770 &   -21.7 &   -48.70$\pm$0.02 &    24.05$\pm$0.02 &    ....\hspace*{0.5cm} &    ....\hspace*{0.5cm}  \\
    537 &              3 &     0.764 &   -30.8 &    61.25$\pm$0.02 &    58.22$\pm$0.02 &    ....\hspace*{0.5cm} &    ....\hspace*{0.5cm}  \\
     75 &        1,2,3,4 &     0.761 &   -28.3 &   -37.42$\pm$0.03 &    36.32$\pm$0.03 &  -24.2$\pm$ 0.5 &    3.4$\pm$ 0.5  \\
     43 &              1 &     0.747 &    -9.6 &    52.56$\pm$0.03 &    14.03$\pm$0.03 &    ....\hspace*{0.5cm} &    ....\hspace*{0.5cm}  \\
    241 &              2 &     0.746 &    -9.4 &    31.69$\pm$0.02 &   -21.77$\pm$0.02 &    ....\hspace*{0.5cm} &    ....\hspace*{0.5cm}  \\
    671 &              4 &     0.695 &     0.6 &    48.94$\pm$0.02 &    -6.73$\pm$0.02 &    ....\hspace*{0.5cm} &    ....\hspace*{0.5cm}  \\
    679 &              4 &     0.693 &    -7.3 &    21.48$\pm$0.02 &   -33.45$\pm$0.02 &    ....\hspace*{0.5cm} &    ....\hspace*{0.5cm}  \\
    706 &              4 &     0.675 &     3.8 &   -52.08$\pm$0.02 &   -60.44$\pm$0.02 &    ....\hspace*{0.5cm} &    ....\hspace*{0.5cm}  \\
     33 &            1,2 &     0.657 &   -15.1 &    22.66$\pm$0.02 &   -28.56$\pm$0.02 &    ....\hspace*{0.5cm} &    ....\hspace*{0.5cm}  \\
     85 &            1,2 &     0.652 &   -32.1 &    59.66$\pm$0.02 &    55.93$\pm$0.02 &    ....\hspace*{0.5cm} &    ....\hspace*{0.5cm}  \\
    273 &          2,3,4 &     0.644 &     0.6 &    53.17$\pm$0.02 &    -1.98$\pm$0.03 &   -6.8$\pm$ 0.7 &  -21.6$\pm$ 0.7  \\
    119 &        1,2,3,4 &     0.642 &   -28.1 &   -44.91$\pm$0.03 &    -1.17$\pm$0.03 &  -24.7$\pm$ 0.4 &   -4.2$\pm$ 0.4  \\
    634 &              4 &     0.637 &    -4.7 &    32.54$\pm$0.02 &   -22.82$\pm$0.02 &    ....\hspace*{0.5cm} &    ....\hspace*{0.5cm}  \\
    635 &              4 &     0.627 &    -4.3 &    32.16$\pm$0.02 &   -24.14$\pm$0.02 &    ....\hspace*{0.5cm} &    ....\hspace*{0.5cm}  \\
    676 &              4 &     0.609 &    -9.1 &    22.86$\pm$0.04 &   -33.55$\pm$0.04 &    ....\hspace*{0.5cm} &    ....\hspace*{0.5cm}  \\
    219 &              2 &     0.608 &   -11.6 &    29.62$\pm$0.02 &   -22.12$\pm$0.02 &    ....\hspace*{0.5cm} &    ....\hspace*{0.5cm}  \\
     29 &        1,2,3,4 &     0.598 &    -7.2 &    19.97$\pm$0.05 &   -33.53$\pm$0.04 &    7.4$\pm$ 0.6 &  -12.8$\pm$ 0.5  \\
    242 &              2 &     0.578 &   -17.1 &    30.66$\pm$0.02 &   -17.56$\pm$0.02 &    ....\hspace*{0.5cm} &    ....\hspace*{0.5cm}  \\
     83 &        1,2,3,4 &     0.570 &   -36.3 &   -41.56$\pm$0.07 &    38.95$\pm$0.06 &  -14.5$\pm$ 0.6 &    9.3$\pm$ 0.6  \\
    408 &              3 &     0.552 &   -13.7 &    27.70$\pm$0.02 &   -22.20$\pm$0.02 &    ....\hspace*{0.5cm} &    ....\hspace*{0.5cm}  \\
    333 &          2,3,4 &     0.548 &     2.6 &   -50.67$\pm$0.02 &   -58.41$\pm$0.02 &  -39.1$\pm$ 0.7 &  -30.2$\pm$ 0.7  \\
    678 &              4 &     0.532 &    -7.4 &    23.28$\pm$0.02 &   -31.77$\pm$0.02 &    ....\hspace*{0.5cm} &    ....\hspace*{0.5cm}  \\
    307 &            2,4 &     0.530 &   -10.0 &    19.88$\pm$0.03 &   -35.34$\pm$0.03 &    ....\hspace*{0.5cm} &    ....\hspace*{0.5cm}  \\
     24 &            1,2 &     0.527 &    -9.3 &    25.95$\pm$0.02 &   -29.14$\pm$0.02 &    ....\hspace*{0.5cm} &    ....\hspace*{0.5cm}  \\
     35 &              1 &     0.517 &   -14.0 &    25.41$\pm$0.03 &   -23.16$\pm$0.03 &    ....\hspace*{0.5cm} &    ....\hspace*{0.5cm}  \\
    520 &              3 &     0.496 &   -18.1 &    16.05$\pm$0.02 &   -33.01$\pm$0.02 &    ....\hspace*{0.5cm} &    ....\hspace*{0.5cm}  \\
    108 &          1,2,3 &     0.482 &    15.1 &   -10.16$\pm$0.02 &   -70.83$\pm$0.02 &   -7.5$\pm$ 0.9 &  -34.4$\pm$ 0.9  \\
    263 &              2 &     0.480 &     0.6 &    49.08$\pm$0.02 &    -6.59$\pm$0.02 &    ....\hspace*{0.5cm} &    ....\hspace*{0.5cm}  \\
    433 &              3 &     0.467 &   -11.5 &    27.66$\pm$0.02 &   -22.50$\pm$0.02 &    ....\hspace*{0.5cm} &    ....\hspace*{0.5cm}  \\
    244 &              2 &     0.466 &    -9.3 &    27.97$\pm$0.02 &   -19.54$\pm$0.03 &    ....\hspace*{0.5cm} &    ....\hspace*{0.5cm}  \\
    522 &              3 &     0.461 &    -6.9 &    23.20$\pm$0.03 &   -31.26$\pm$0.03 &    ....\hspace*{0.5cm} &    ....\hspace*{0.5cm}  \\
     20 &              1 &     0.456 &    -8.1 &    29.92$\pm$0.03 &   -21.13$\pm$0.03 &    ....\hspace*{0.5cm} &    ....\hspace*{0.5cm}  \\
    689 &              4 &     0.450 &   -17.8 &   -52.27$\pm$0.03 &    27.92$\pm$0.03 &    ....\hspace*{0.5cm} &    ....\hspace*{0.5cm}  \\
    624 &              4 &     0.447 &   -17.7 &    27.69$\pm$0.03 &   -23.94$\pm$0.03 &    ....\hspace*{0.5cm} &    ....\hspace*{0.5cm}  \\
    122 &              1 &     0.447 &   -27.1 &   -36.36$\pm$0.03 &    -1.26$\pm$0.03 &    ....\hspace*{0.5cm} &    ....\hspace*{0.5cm}  \\
    125 &              1 &     0.446 &     9.3 &   -87.22$\pm$0.02 &   -96.55$\pm$0.02 &    ....\hspace*{0.5cm} &    ....\hspace*{0.5cm}  \\
     22 &          1,2,3 &     0.446 &    -7.5 &    25.98$\pm$0.02 &   -30.24$\pm$0.02 &   -1.4$\pm$ 0.6 &  -20.7$\pm$ 0.6  \\
    275 &          2,3,4 &     0.425 &     1.3 &    51.28$\pm$0.04 &    -5.39$\pm$0.04 &   -4.5$\pm$ 0.8 &  -11.8$\pm$ 0.9  \\
    440 &              3 &     0.422 &    -2.1 &    30.48$\pm$0.02 &   -27.17$\pm$0.02 &    ....\hspace*{0.5cm} &    ....\hspace*{0.5cm}  \\
    124 &        1,2,3,4 &     0.410 &   -27.6 &   -38.71$\pm$0.03 &     1.37$\pm$0.03 &   -4.5$\pm$ 0.5 &   11.2$\pm$ 0.4  \\
     48 &            1,2 &     0.409 &    -2.6 &    54.38$\pm$0.02 &     4.93$\pm$0.02 &    ....\hspace*{0.5cm} &    ....\hspace*{0.5cm}  \\
     47 &          1,2,3 &     0.406 &    -3.6 &    55.12$\pm$0.03 &     6.52$\pm$0.03 &    3.0$\pm$ 0.7 &   -5.9$\pm$ 0.7  \\
     70 &              1 &     0.404 &   -17.4 &     1.34$\pm$0.03 &   -39.98$\pm$0.03 &    ....\hspace*{0.5cm} &    ....\hspace*{0.5cm}  \\
     95 &              1 &     0.403 &    -2.1 &   -49.99$\pm$0.03 &   -55.38$\pm$0.02 &    ....\hspace*{0.5cm} &    ....\hspace*{0.5cm}  \\
     77 &        1,2,3,4 &     0.396 &   -32.4 &   -38.37$\pm$0.03 &    31.69$\pm$0.03 &  -18.9$\pm$ 0.6 &   18.5$\pm$ 0.7  \\
    682 &              4 &     0.388 &    -7.2 &    25.05$\pm$0.03 &   -29.84$\pm$0.03 &    ....\hspace*{0.5cm} &    ....\hspace*{0.5cm}  \\
    247 &            2,3 &     0.386 &    -8.1 &    26.54$\pm$0.03 &   -29.47$\pm$0.03 &    ....\hspace*{0.5cm} &    ....\hspace*{0.5cm}  \\
     94 &          1,2,3 &     0.381 &    -1.5 &   -49.34$\pm$0.03 &   -55.43$\pm$0.03 &  -21.2$\pm$ 0.7 &  -28.7$\pm$ 0.7  \\
    329 &          2,3,4 &     0.341 &   -34.0 &    60.38$\pm$0.03 &    54.96$\pm$0.03 &   24.4$\pm$ 0.8 &   31.6$\pm$ 0.8  \\
    225 &              2 &     0.331 &    -2.7 &    28.51$\pm$0.02 &   -19.00$\pm$0.02 &    ....\hspace*{0.5cm} &    ....\hspace*{0.5cm}  \\
    402 &              3 &     0.328 &   -10.7 &    30.15$\pm$0.02 &   -20.74$\pm$0.02 &    ....\hspace*{0.5cm} &    ....\hspace*{0.5cm}  \\
    115 &              1 &     0.323 &   -14.8 &   -51.56$\pm$0.03 &    11.30$\pm$0.03 &    ....\hspace*{0.5cm} &    ....\hspace*{0.5cm}  \\
     46 &              1 &     0.318 &    -9.8 &    52.64$\pm$0.05 &    11.70$\pm$0.05 &    ....\hspace*{0.5cm} &    ....\hspace*{0.5cm}  \\
    116 &              1 &     0.318 &   -14.5 &   -51.85$\pm$0.03 &    10.23$\pm$0.06 &    ....\hspace*{0.5cm} &    ....\hspace*{0.5cm}  \\
    477 &              3 &     0.316 &   -11.6 &    53.31$\pm$0.02 &    12.17$\pm$0.03 &    ....\hspace*{0.5cm} &    ....\hspace*{0.5cm}  \\
     36 &              1 &     0.316 &    -9.0 &    29.39$\pm$0.04 &   -24.77$\pm$0.03 &    ....\hspace*{0.5cm} &    ....\hspace*{0.5cm}  \\
    722 &              4 &     0.310 &    -3.8 &   -89.32$\pm$0.03 &   -83.91$\pm$0.03 &    ....\hspace*{0.5cm} &    ....\hspace*{0.5cm}  \\
    117 &            1,2 &     0.310 &   -18.4 &   -36.75$\pm$0.02 &    -4.99$\pm$0.02 &    ....\hspace*{0.5cm} &    ....\hspace*{0.5cm}  \\
    533 &              3 &     0.302 &     4.0 &   -51.57$\pm$0.02 &   -59.99$\pm$0.02 &    ....\hspace*{0.5cm} &    ....\hspace*{0.5cm}  \\
    479 &              3 &     0.295 &   -28.0 &    44.17$\pm$0.02 &    15.49$\pm$0.02 &    ....\hspace*{0.5cm} &    ....\hspace*{0.5cm}  \\
    545 &              3 &     0.295 &    -7.5 &    11.36$\pm$0.02 &   -38.40$\pm$0.02 &    ....\hspace*{0.5cm} &    ....\hspace*{0.5cm}  \\
     21 &          1,2,3 &     0.291 &    -7.4 &    26.96$\pm$0.02 &   -28.96$\pm$0.02 &   11.4$\pm$ 0.7 &   -6.6$\pm$ 0.8  \\
    419 &              3 &     0.288 &   -11.0 &    31.12$\pm$0.02 &   -22.24$\pm$0.03 &    ....\hspace*{0.5cm} &    ....\hspace*{0.5cm}  \\
    627 &              4 &     0.280 &   -18.1 &    28.26$\pm$0.02 &   -18.75$\pm$0.03 &    ....\hspace*{0.5cm} &    ....\hspace*{0.5cm}  \\
     49 &              1 &     0.275 &    -9.1 &    52.01$\pm$0.03 &    18.99$\pm$0.03 &    ....\hspace*{0.5cm} &    ....\hspace*{0.5cm}  \\
     27 &        1,2,3,4 &     0.274 &    -7.8 &    23.51$\pm$0.02 &   -31.72$\pm$0.02 &    3.6$\pm$ 0.4 &  -19.0$\pm$ 0.4  \\
    525 &            3,4 &     0.269 &    -3.0 &    13.42$\pm$0.03 &   -37.41$\pm$0.03 &    ....\hspace*{0.5cm} &    ....\hspace*{0.5cm}  \\
    109 &              1 &     0.266 &    15.1 &   -10.55$\pm$0.03 &   -71.88$\pm$0.02 &    ....\hspace*{0.5cm} &    ....\hspace*{0.5cm}  \\
    248 &              2 &     0.256 &    -4.2 &    29.69$\pm$0.02 &   -16.87$\pm$0.03 &    ....\hspace*{0.5cm} &    ....\hspace*{0.5cm}  \\
     25 &            1,2 &     0.255 &    -8.4 &    25.06$\pm$0.02 &   -30.57$\pm$0.02 &    ....\hspace*{0.5cm} &    ....\hspace*{0.5cm}  \\
     88 &              1 &     0.248 &   -33.8 &    61.31$\pm$0.03 &    51.37$\pm$0.03 &    ....\hspace*{0.5cm} &    ....\hspace*{0.5cm}  \\
     18 &              1 &     0.247 &   -22.5 &    26.83$\pm$0.02 &   -19.56$\pm$0.02 &    ....\hspace*{0.5cm} &    ....\hspace*{0.5cm}  \\
     66 &            1,2 &     0.238 &    -5.1 &   -79.33$\pm$0.03 &   -74.31$\pm$0.03 &    ....\hspace*{0.5cm} &    ....\hspace*{0.5cm}  \\
    110 &            1,2 &     0.237 &    15.6 &   -10.50$\pm$0.02 &   -72.63$\pm$0.02 &    ....\hspace*{0.5cm} &    ....\hspace*{0.5cm}  \\
    611 &              4 &     0.234 &   -11.5 &    25.20$\pm$0.04 &   -28.08$\pm$0.03 &    ....\hspace*{0.5cm} &    ....\hspace*{0.5cm}  \\
     63 &          1,2,3 &     0.231 &    -1.2 &    42.52$\pm$0.06 &    -5.74$\pm$0.05 &   -2.7$\pm$ 1.1 &  -15.4$\pm$ 1.1  \\
    484 &              3 &     0.229 &   -11.5 &    54.99$\pm$0.03 &    16.55$\pm$0.03 &    ....\hspace*{0.5cm} &    ....\hspace*{0.5cm}  \\
     81 &        1,2,3,4 &     0.229 &   -26.6 &   -44.66$\pm$0.04 &    34.80$\pm$0.04 &  -18.0$\pm$ 0.5 &    6.3$\pm$ 0.6  \\
    444 &              3 &     0.228 &   -10.2 &    28.17$\pm$0.03 &   -19.57$\pm$0.03 &    ....\hspace*{0.5cm} &    ....\hspace*{0.5cm}  \\
    603 &              4 &     0.221 &    -7.6 &    31.50$\pm$0.03 &   -19.54$\pm$0.03 &    ....\hspace*{0.5cm} &    ....\hspace*{0.5cm}  \\
    475 &            3,4 &     0.218 &     0.5 &    54.58$\pm$0.05 &    -0.93$\pm$0.06 &    ....\hspace*{0.5cm} &    ....\hspace*{0.5cm}  \\
    686 &              4 &     0.214 &   -10.6 &    26.49$\pm$0.03 &   -28.59$\pm$0.03 &    ....\hspace*{0.5cm} &    ....\hspace*{0.5cm}  \\
     61 &              1 &     0.212 &    -4.4 &    52.92$\pm$0.03 &     1.90$\pm$0.03 &    ....\hspace*{0.5cm} &    ....\hspace*{0.5cm}  \\
    495 &            3,4 &     0.211 &   -17.6 &   -52.10$\pm$0.03 &    25.96$\pm$0.04 &    ....\hspace*{0.5cm} &    ....\hspace*{0.5cm}  \\
    523 &              3 &     0.207 &    -7.7 &    24.06$\pm$0.02 &   -30.46$\pm$0.03 &    ....\hspace*{0.5cm} &    ....\hspace*{0.5cm}  \\
    521 &              3 &     0.205 &   -17.5 &    15.49$\pm$0.03 &   -33.46$\pm$0.03 &    ....\hspace*{0.5cm} &    ....\hspace*{0.5cm}  \\
     30 &              1 &     0.204 &    -6.4 &    17.78$\pm$0.03 &   -34.98$\pm$0.03 &    ....\hspace*{0.5cm} &    ....\hspace*{0.5cm}  \\
    292 &              2 &     0.203 &   -15.1 &   -52.35$\pm$0.03 &    11.03$\pm$0.05 &    ....\hspace*{0.5cm} &    ....\hspace*{0.5cm}  \\
    547 &            3,4 &     0.197 &     8.8 &   -89.98$\pm$0.02 &   -99.98$\pm$0.02 &    ....\hspace*{0.5cm} &    ....\hspace*{0.5cm}  \\
    452 &              3 &     0.197 &    -4.8 &    51.64$\pm$0.03 &    -0.39$\pm$0.03 &    ....\hspace*{0.5cm} &    ....\hspace*{0.5cm}  \\
    126 &        1,2,3,4 &     0.196 &     1.9 &   -93.18$\pm$0.04 &   -94.71$\pm$0.04 &  -27.0$\pm$ 0.5 &  -29.5$\pm$ 0.5  \\
    656 &              4 &     0.193 &   -42.5 &   -36.59$\pm$0.03 &    32.42$\pm$0.03 &    ....\hspace*{0.5cm} &    ....\hspace*{0.5cm}  \\
    320 &            2,3 &     0.189 &   -27.7 &   -44.49$\pm$0.02 &    35.83$\pm$0.03 &    ....\hspace*{0.5cm} &    ....\hspace*{0.5cm}  \\
     51 &              1 &     0.189 &   -18.7 &    47.77$\pm$0.03 &     5.35$\pm$0.03 &    ....\hspace*{0.5cm} &    ....\hspace*{0.5cm}  \\
    654 &              4 &     0.187 &   -35.5 &   -38.28$\pm$0.03 &    33.66$\pm$0.03 &    ....\hspace*{0.5cm} &    ....\hspace*{0.5cm}  \\
    103 &        1,2,3,4 &     0.179 &   -20.1 &   -49.70$\pm$0.03 &    18.69$\pm$0.03 &  -19.4$\pm$ 0.4 &  -18.1$\pm$ 0.5  \\
    684 &              4 &     0.178 &   -20.6 &    13.71$\pm$0.03 &   -35.76$\pm$0.03 &    ....\hspace*{0.5cm} &    ....\hspace*{0.5cm}  \\
    712 &              4 &     0.150 &   -24.2 &   -48.08$\pm$0.03 &    -2.76$\pm$0.03 &    ....\hspace*{0.5cm} &    ....\hspace*{0.5cm}  \\
    252 &            2,3 &     0.149 &   -17.5 &    26.32$\pm$0.02 &   -22.86$\pm$0.03 &    ....\hspace*{0.5cm} &    ....\hspace*{0.5cm}  \\
    462 &              3 &     0.146 &     0.8 &    48.93$\pm$0.02 &    -6.87$\pm$0.02 &    ....\hspace*{0.5cm} &    ....\hspace*{0.5cm}  \\
     32 &              1 &     0.145 &    -7.1 &    16.55$\pm$0.03 &   -35.78$\pm$0.03 &    ....\hspace*{0.5cm} &    ....\hspace*{0.5cm}  \\
    488 &              3 &     0.145 &    -4.6 &   -79.09$\pm$0.03 &   -72.74$\pm$0.03 &    ....\hspace*{0.5cm} &    ....\hspace*{0.5cm}  \\
    206 &              2 &     0.145 &   -17.6 &    29.50$\pm$0.03 &   -21.16$\pm$0.04 &    ....\hspace*{0.5cm} &    ....\hspace*{0.5cm}  \\
    306 &            2,3 &     0.141 &   -17.2 &    20.87$\pm$0.03 &   -31.65$\pm$0.03 &    ....\hspace*{0.5cm} &    ....\hspace*{0.5cm}  \\
     79 &              1 &     0.134 &   -32.9 &   -35.74$\pm$0.03 &    29.31$\pm$0.03 &    ....\hspace*{0.5cm} &    ....\hspace*{0.5cm}  \\
    614 &              4 &     0.130 &   -20.2 &    28.39$\pm$0.03 &   -21.19$\pm$0.03 &    ....\hspace*{0.5cm} &    ....\hspace*{0.5cm}  \\
    463 &              3 &     0.129 &     0.8 &    48.64$\pm$0.03 &    -7.24$\pm$0.03 &    ....\hspace*{0.5cm} &    ....\hspace*{0.5cm}  \\
    123 &            1,2 &     0.129 &   -24.6 &   -38.39$\pm$0.03 &     4.28$\pm$0.03 &    ....\hspace*{0.5cm} &    ....\hspace*{0.5cm}  \\
    528 &              3 &     0.128 &   -17.3 &    22.47$\pm$0.04 &   -30.67$\pm$0.04 &    ....\hspace*{0.5cm} &    ....\hspace*{0.5cm}  \\
    492 &              3 &     0.126 &   -23.8 &   -14.81$\pm$0.03 &   -22.03$\pm$0.03 &    ....\hspace*{0.5cm} &    ....\hspace*{0.5cm}  \\
     96 &            1,2 &     0.126 &    -2.3 &   -51.68$\pm$0.04 &   -55.52$\pm$0.03 &    ....\hspace*{0.5cm} &    ....\hspace*{0.5cm}  \\
    450 &            3,4 &     0.126 &    -0.5 &    28.89$\pm$0.08 &   -28.80$\pm$0.10 &    ....\hspace*{0.5cm} &    ....\hspace*{0.5cm}  \\
    456 &              3 &     0.123 &    -4.9 &    52.87$\pm$0.03 &     1.27$\pm$0.02 &    ....\hspace*{0.5cm} &    ....\hspace*{0.5cm}  \\
     92 &              1 &     0.121 &     3.3 &   -49.13$\pm$0.03 &   -57.78$\pm$0.03 &    ....\hspace*{0.5cm} &    ....\hspace*{0.5cm}  \\
    693 &              4 &     0.121 &   -24.9 &   -46.59$\pm$0.04 &    14.41$\pm$0.04 &    ....\hspace*{0.5cm} &    ....\hspace*{0.5cm}  \\
    111 &          1,2,3 &     0.120 &    11.0 &    -9.22$\pm$0.03 &   -64.87$\pm$0.03 &  -14.6$\pm$ 0.8 &  -30.3$\pm$ 1.0  \\
    323 &          2,3,4 &     0.120 &   -34.0 &   -37.11$\pm$0.05 &    33.84$\pm$0.07 &  -24.1$\pm$ 1.1 &    9.9$\pm$ 1.4  \\
    536 &            3,4 &     0.118 &     4.0 &   -50.59$\pm$0.03 &   -59.49$\pm$0.03 &    ....\hspace*{0.5cm} &    ....\hspace*{0.5cm}  \\
    541 &              3 &     0.118 &   -32.3 &    60.89$\pm$0.03 &    56.24$\pm$0.03 &    ....\hspace*{0.5cm} &    ....\hspace*{0.5cm}  \\
    608 &              4 &     0.117 &    -6.6 &    32.50$\pm$0.04 &   -21.98$\pm$0.03 &    ....\hspace*{0.5cm} &    ....\hspace*{0.5cm}  \\
    348 &            2,3 &     0.113 &   -27.0 &   -27.50$\pm$0.02 &   -17.80$\pm$0.03 &    ....\hspace*{0.5cm} &    ....\hspace*{0.5cm}  \\
    471 &              3 &     0.112 &     1.1 &    51.86$\pm$0.03 &    -5.10$\pm$0.03 &    ....\hspace*{0.5cm} &    ....\hspace*{0.5cm}  \\
    350 &          2,3,4 &     0.110 &     9.3 &   -91.73$\pm$0.03 &  -104.87$\pm$0.04 &  -26.2$\pm$ 0.9 &  -30.7$\pm$ 1.0  \\
    723 &              4 &     0.108 &     5.8 &   -28.85$\pm$0.04 &   -57.31$\pm$0.04 &    ....\hspace*{0.5cm} &    ....\hspace*{0.5cm}  \\
    457 &              3 &     0.108 &    -4.9 &    53.21$\pm$0.04 &     1.61$\pm$0.03 &    ....\hspace*{0.5cm} &    ....\hspace*{0.5cm}  \\
    647 &              4 &     0.107 &   -31.9 &    61.64$\pm$0.03 &    57.26$\pm$0.03 &    ....\hspace*{0.5cm} &    ....\hspace*{0.5cm}  \\
    500 &            3,4 &     0.105 &   -18.5 &   -51.43$\pm$0.06 &    26.65$\pm$0.08 &    ....\hspace*{0.5cm} &    ....\hspace*{0.5cm}  \\
    253 &              2 &     0.105 &    -6.8 &    27.93$\pm$0.03 &   -28.34$\pm$0.03 &    ....\hspace*{0.5cm} &    ....\hspace*{0.5cm}  \\
    259 &              2 &     0.103 &    -1.9 &    51.68$\pm$0.04 &    -0.41$\pm$0.04 &    ....\hspace*{0.5cm} &    ....\hspace*{0.5cm}  \\
    129 &              1 &     0.103 &    -6.6 &    77.31$\pm$0.05 &    18.09$\pm$0.04 &    ....\hspace*{0.5cm} &    ....\hspace*{0.5cm}  \\
    438 &              3 &     0.102 &    -8.9 &    25.89$\pm$0.05 &   -30.55$\pm$0.04 &    ....\hspace*{0.5cm} &    ....\hspace*{0.5cm}  \\
    711 &              4 &     0.101 &   -25.6 &   -47.72$\pm$0.04 &    -1.99$\pm$0.04 &    ....\hspace*{0.5cm} &    ....\hspace*{0.5cm}  \\
     38 &          1,2,3 &     0.100 &    -1.3 &    26.82$\pm$0.03 &   -23.43$\pm$0.03 &   16.1$\pm$ 0.9 &    0.8$\pm$ 0.9  \\
    546 &              3 &     0.100 &    -9.3 &    11.16$\pm$0.03 &   -38.39$\pm$0.03 &    ....\hspace*{0.5cm} &    ....\hspace*{0.5cm}  \\
    102 &        1,2,3,4 &     0.099 &   -23.1 &   -48.14$\pm$0.04 &    25.59$\pm$0.04 &  -16.4$\pm$ 0.7 &   -1.0$\pm$ 0.8  \\
    724 &              4 &     0.097 &   -39.8 &   123.24$\pm$0.04 &   201.93$\pm$0.04 &    ....\hspace*{0.5cm} &    ....\hspace*{0.5cm}  \\
    695 &              4 &     0.096 &   -18.5 &   -52.63$\pm$0.04 &    24.36$\pm$0.06 &    ....\hspace*{0.5cm} &    ....\hspace*{0.5cm}  \\
    276 &          2,3,4 &     0.095 &    -1.8 &    53.52$\pm$0.04 &     0.47$\pm$0.05 &   -4.0$\pm$ 1.2 &  -12.7$\pm$ 1.2  \\
    636 &              4 &     0.094 &   -18.0 &    32.63$\pm$0.04 &   -15.52$\pm$0.04 &    ....\hspace*{0.5cm} &    ....\hspace*{0.5cm}  \\
    216 &              2 &     0.092 &    -8.8 &    27.42$\pm$0.03 &   -18.26$\pm$0.04 &    ....\hspace*{0.5cm} &    ....\hspace*{0.5cm}  \\
    112 &              1 &     0.090 &    11.9 &    -9.09$\pm$0.04 &   -66.03$\pm$0.05 &    ....\hspace*{0.5cm} &    ....\hspace*{0.5cm}  \\
    268 &            2,3 &     0.090 &    -2.2 &    50.28$\pm$0.02 &    -2.91$\pm$0.03 &    ....\hspace*{0.5cm} &    ....\hspace*{0.5cm}  \\
    658 &              4 &     0.088 &   -25.4 &   -38.09$\pm$0.04 &    44.29$\pm$0.04 &    ....\hspace*{0.5cm} &    ....\hspace*{0.5cm}  \\
    271 &              2 &     0.086 &   -18.5 &    45.93$\pm$0.03 &     2.49$\pm$0.03 &    ....\hspace*{0.5cm} &    ....\hspace*{0.5cm}  \\
    332 &              2 &     0.085 &     1.9 &   -51.58$\pm$0.03 &   -58.25$\pm$0.03 &    ....\hspace*{0.5cm} &    ....\hspace*{0.5cm}  \\
    127 &            1,2 &     0.085 &   -38.8 &    53.99$\pm$0.03 &    41.02$\pm$0.03 &    ....\hspace*{0.5cm} &    ....\hspace*{0.5cm}  \\
    685 &              4 &     0.085 &    -7.8 &    26.08$\pm$0.07 &   -31.62$\pm$0.06 &    ....\hspace*{0.5cm} &    ....\hspace*{0.5cm}  \\
     87 &              1 &     0.084 &   -34.6 &    61.11$\pm$0.04 &    59.06$\pm$0.03 &    ....\hspace*{0.5cm} &    ....\hspace*{0.5cm}  \\
     31 &            1,2 &     0.083 &    -7.5 &    18.53$\pm$0.06 &   -34.80$\pm$0.04 &    ....\hspace*{0.5cm} &    ....\hspace*{0.5cm}  \\
    251 &          2,3,4 &     0.083 &     2.0 &    25.22$\pm$0.02 &   -25.07$\pm$0.03 &   22.4$\pm$ 1.3 &    8.4$\pm$ 1.4  \\
     53 &              1 &     0.081 &   -21.7 &    51.39$\pm$0.04 &    10.14$\pm$0.04 &    ....\hspace*{0.5cm} &    ....\hspace*{0.5cm}  \\
     62 &          1,2,3 &     0.080 &     0.3 &    41.69$\pm$0.04 &    -6.59$\pm$0.04 &   -3.1$\pm$ 1.3 &   -8.1$\pm$ 1.2  \\
    281 &              2 &     0.080 &   -32.2 &    42.98$\pm$0.03 &    15.28$\pm$0.03 &    ....\hspace*{0.5cm} &    ....\hspace*{0.5cm}  \\
    294 &          2,3,4 &     0.079 &   -18.6 &   -51.42$\pm$0.04 &    24.16$\pm$0.04 &  -25.9$\pm$ 1.1 &  -11.6$\pm$ 1.3  \\
    289 &            2,3 &     0.078 &    -8.7 &    54.33$\pm$0.03 &    10.67$\pm$0.04 &    ....\hspace*{0.5cm} &    ....\hspace*{0.5cm}  \\
    312 &              2 &     0.078 &    -9.0 &    22.64$\pm$0.04 &   -32.86$\pm$0.05 &    ....\hspace*{0.5cm} &    ....\hspace*{0.5cm}  \\
    130 &              1 &     0.077 &    -4.3 &   117.92$\pm$0.07 &    -4.21$\pm$0.07 &    ....\hspace*{0.5cm} &    ....\hspace*{0.5cm}  \\
    720 &              4 &     0.076 &    -0.3 &   -96.91$\pm$0.05 &   -96.55$\pm$0.05 &    ....\hspace*{0.5cm} &    ....\hspace*{0.5cm}  \\
    702 &              4 &     0.076 &    -4.5 &   -79.58$\pm$0.06 &   -72.55$\pm$0.06 &    ....\hspace*{0.5cm} &    ....\hspace*{0.5cm}  \\
    313 &              2 &     0.075 &   -22.5 &    12.15$\pm$0.03 &   -35.52$\pm$0.03 &    ....\hspace*{0.5cm} &    ....\hspace*{0.5cm}  \\
    106 &              1 &     0.075 &   -25.3 &   -47.06$\pm$0.05 &    25.83$\pm$0.06 &    ....\hspace*{0.5cm} &    ....\hspace*{0.5cm}  \\
    491 &              3 &     0.074 &   -32.7 &   -16.93$\pm$0.03 &   -25.92$\pm$0.03 &    ....\hspace*{0.5cm} &    ....\hspace*{0.5cm}  \\
    354 &              2 &     0.074 &   -40.7 &   123.37$\pm$0.03 &   199.37$\pm$0.04 &    ....\hspace*{0.5cm} &    ....\hspace*{0.5cm}  \\
    531 &              3 &     0.072 &    -2.4 &   -50.73$\pm$0.03 &   -56.32$\pm$0.04 &    ....\hspace*{0.5cm} &    ....\hspace*{0.5cm}  \\
     71 &              1 &     0.070 &   -19.0 &    11.46$\pm$0.04 &   -35.70$\pm$0.04 &    ....\hspace*{0.5cm} &    ....\hspace*{0.5cm}  \\
    725 &              4 &     0.069 &   -28.9 &   -27.28$\pm$0.05 &   -29.29$\pm$0.05 &    ....\hspace*{0.5cm} &    ....\hspace*{0.5cm}  \\
     99 &            1,2 &     0.068 &   -21.2 &   -49.21$\pm$0.04 &    23.07$\pm$0.03 &    ....\hspace*{0.5cm} &    ....\hspace*{0.5cm}  \\
    554 &            3,4 &     0.067 &   -30.0 &   -25.81$\pm$0.03 &   -30.06$\pm$0.03 &    ....\hspace*{0.5cm} &    ....\hspace*{0.5cm}  \\
    687 &              4 &     0.066 &   -20.8 &    15.87$\pm$0.05 &   -33.12$\pm$0.05 &    ....\hspace*{0.5cm} &    ....\hspace*{0.5cm}  \\
    728 &              4 &     0.065 &    11.5 &   -10.60$\pm$0.05 &   -66.92$\pm$0.06 &    ....\hspace*{0.5cm} &    ....\hspace*{0.5cm}  \\
    727 &              4 &     0.065 &   -36.2 &   -12.23$\pm$0.05 &   -14.68$\pm$0.05 &    ....\hspace*{0.5cm} &    ....\hspace*{0.5cm}  \\
    128 &            1,2 &     0.064 &   -38.7 &    54.15$\pm$0.04 &    42.04$\pm$0.04 &    ....\hspace*{0.5cm} &    ....\hspace*{0.5cm}  \\
    672 &              4 &     0.063 &    -5.6 &    51.13$\pm$0.06 &     2.73$\pm$0.06 &    ....\hspace*{0.5cm} &    ....\hspace*{0.5cm}  \\
    552 &              3 &     0.063 &    -0.7 &    41.15$\pm$0.04 &    -6.67$\pm$0.04 &    ....\hspace*{0.5cm} &    ....\hspace*{0.5cm}  \\
    730 &              4 &     0.062 &     0.8 &   -69.15$\pm$0.05 &   -70.12$\pm$0.05 &    ....\hspace*{0.5cm} &    ....\hspace*{0.5cm}  \\
    703 &              4 &     0.062 &    -4.6 &   -81.99$\pm$0.04 &   -74.84$\pm$0.05 &    ....\hspace*{0.5cm} &    ....\hspace*{0.5cm}  \\
    709 &              4 &     0.062 &    -0.6 &   -51.31$\pm$0.06 &   -58.21$\pm$0.05 &    ....\hspace*{0.5cm} &    ....\hspace*{0.5cm}  \\
    322 &          2,3,4 &     0.062 &   -30.4 &   -39.05$\pm$0.06 &    34.30$\pm$0.09 &  -16.2$\pm$ 1.5 &   -1.0$\pm$ 1.9  \\
    694 &              4 &     0.061 &   -26.4 &   -47.11$\pm$0.07 &    13.17$\pm$0.07 &    ....\hspace*{0.5cm} &    ....\hspace*{0.5cm}  \\
    101 &        1,2,3,4 &     0.059 &   -20.4 &   -49.46$\pm$0.08 &    20.59$\pm$0.08 &  -23.6$\pm$ 1.0 &  -16.7$\pm$ 1.1  \\
     72 &            1,2 &     0.058 &    -1.1 &    10.60$\pm$0.05 &   -38.34$\pm$0.04 &    ....\hspace*{0.5cm} &    ....\hspace*{0.5cm}  \\
     50 &            1,2 &     0.057 &    -7.3 &    53.83$\pm$0.07 &     8.76$\pm$0.06 &    ....\hspace*{0.5cm} &    ....\hspace*{0.5cm}  \\
    466 &            3,4 &     0.056 &   -19.1 &    47.70$\pm$0.04 &     4.83$\pm$0.05 &    ....\hspace*{0.5cm} &    ....\hspace*{0.5cm}  \\
    120 &            1,2 &     0.055 &   -27.7 &   -45.36$\pm$0.08 &    -2.44$\pm$0.08 &    ....\hspace*{0.5cm} &    ....\hspace*{0.5cm}  \\
     78 &            1,2 &     0.054 &   -32.8 &   -37.79$\pm$0.07 &    32.49$\pm$0.07 &    ....\hspace*{0.5cm} &    ....\hspace*{0.5cm}  \\
    535 &              3 &     0.054 &     2.6 &   -52.62$\pm$0.04 &   -59.41$\pm$0.05 &    ....\hspace*{0.5cm} &    ....\hspace*{0.5cm}  \\
    325 &            2,3 &     0.052 &   -28.2 &   -25.17$\pm$0.04 &   -28.27$\pm$0.05 &    ....\hspace*{0.5cm} &    ....\hspace*{0.5cm}  \\
     90 &              1 &     0.051 &   -33.3 &   -24.56$\pm$0.05 &   -30.49$\pm$0.05 &    ....\hspace*{0.5cm} &    ....\hspace*{0.5cm}  \\
    255 &              2 &     0.051 &     2.1 &    22.65$\pm$0.04 &   -27.97$\pm$0.05 &    ....\hspace*{0.5cm} &    ....\hspace*{0.5cm}  \\
    715 &              4 &     0.050 &   -21.0 &   -37.58$\pm$0.07 &    -5.39$\pm$0.07 &    ....\hspace*{0.5cm} &    ....\hspace*{0.5cm}  \\
     80 &              1 &     0.050 &   -32.4 &   -36.42$\pm$0.07 &    33.98$\pm$0.06 &    ....\hspace*{0.5cm} &    ....\hspace*{0.5cm}  \\
     54 &            1,2 &     0.050 &   -21.7 &    43.42$\pm$0.05 &     6.90$\pm$0.04 &    ....\hspace*{0.5cm} &    ....\hspace*{0.5cm}  \\
    357 &              2 &     0.049 &    -8.0 &   103.62$\pm$0.04 &    67.28$\pm$0.04 &    ....\hspace*{0.5cm} &    ....\hspace*{0.5cm}  \\
    721 &              4 &     0.048 &     4.9 &   -91.97$\pm$0.06 &   -98.36$\pm$0.06 &    ....\hspace*{0.5cm} &    ....\hspace*{0.5cm}  \\
    256 &              2 &     0.048 &   -23.9 &    26.28$\pm$0.03 &   -19.84$\pm$0.05 &    ....\hspace*{0.5cm} &    ....\hspace*{0.5cm}  \\
    107 &            1,2 &     0.047 &   -17.9 &   -50.85$\pm$0.07 &    17.76$\pm$0.07 &    ....\hspace*{0.5cm} &    ....\hspace*{0.5cm}  \\
    330 &          2,3,4 &     0.046 &   -29.7 &    59.95$\pm$0.04 &    58.83$\pm$0.07 &   26.2$\pm$ 1.6 &   29.2$\pm$ 1.9  \\
    515 &              3 &     0.046 &   -31.3 &   -38.75$\pm$0.06 &    33.90$\pm$0.07 &    ....\hspace*{0.5cm} &    ....\hspace*{0.5cm}  \\
    507 &              3 &     0.045 &   -28.8 &   -38.92$\pm$0.07 &    35.98$\pm$0.05 &    ....\hspace*{0.5cm} &    ....\hspace*{0.5cm}  \\
    346 &              2 &     0.044 &   -26.7 &   -46.32$\pm$0.04 &    -1.52$\pm$0.06 &    ....\hspace*{0.5cm} &    ....\hspace*{0.5cm}  \\
    358 &              2 &     0.043 &   -19.8 &   -10.34$\pm$0.04 &   -39.71$\pm$0.05 &    ....\hspace*{0.5cm} &    ....\hspace*{0.5cm}  \\
    131 &              1 &     0.043 &   -40.7 &   123.03$\pm$0.06 &   198.14$\pm$0.07 &    ....\hspace*{0.5cm} &    ....\hspace*{0.5cm}  \\
    544 &              3 &     0.042 &   -29.5 &   -45.37$\pm$0.05 &     4.55$\pm$0.06 &    ....\hspace*{0.5cm} &    ....\hspace*{0.5cm}  \\
    319 &              2 &     0.042 &   -26.7 &   -45.55$\pm$0.04 &    34.28$\pm$0.06 &    ....\hspace*{0.5cm} &    ....\hspace*{0.5cm}  \\
    731 &              4 &     0.042 &   -41.4 &    28.53$\pm$0.07 &    62.00$\pm$0.06 &    ....\hspace*{0.5cm} &    ....\hspace*{0.5cm}  \\
    100 &        1,2,3,4 &     0.041 &   -20.1 &   -49.38$\pm$0.05 &    21.89$\pm$0.06 &  -29.0$\pm$ 1.2 &   -8.7$\pm$ 1.4  \\
    132 &              1 &     0.041 &     5.2 &   -27.65$\pm$0.08 &   -55.26$\pm$0.07 &    ....\hspace*{0.5cm} &    ....\hspace*{0.5cm}  \\
     82 &              1 &     0.041 &   -27.4 &   -42.60$\pm$0.07 &    34.82$\pm$0.06 &    ....\hspace*{0.5cm} &    ....\hspace*{0.5cm}  \\
    345 &              2 &     0.041 &   -25.9 &   -37.65$\pm$0.05 &    -0.06$\pm$0.06 &    ....\hspace*{0.5cm} &    ....\hspace*{0.5cm}  \\
    673 &              4 &     0.040 &     0.4 &    54.63$\pm$0.08 &     0.64$\pm$0.06 &    ....\hspace*{0.5cm} &    ....\hspace*{0.5cm}  \\
    118 &              1 &     0.039 &   -20.4 &   -36.16$\pm$0.10 &    -4.23$\pm$0.08 &    ....\hspace*{0.5cm} &    ....\hspace*{0.5cm}  \\
     67 &              1 &     0.039 &     2.6 &   -74.85$\pm$0.06 &   -75.99$\pm$0.06 &    ....\hspace*{0.5cm} &    ....\hspace*{0.5cm}  \\
    104 &              1 &     0.038 &   -20.0 &   -49.44$\pm$0.08 &    17.18$\pm$0.07 &    ....\hspace*{0.5cm} &    ....\hspace*{0.5cm}  \\
    729 &              4 &     0.037 &    11.0 &   -11.29$\pm$0.09 &   -65.44$\pm$0.08 &    ....\hspace*{0.5cm} &    ....\hspace*{0.5cm}  \\
    550 &              3 &     0.036 &     1.5 &   -94.61$\pm$0.08 &   -97.11$\pm$0.08 &    ....\hspace*{0.5cm} &    ....\hspace*{0.5cm}  \\
    501 &              3 &     0.036 &   -26.5 &   -45.73$\pm$0.06 &    13.31$\pm$0.07 &    ....\hspace*{0.5cm} &    ....\hspace*{0.5cm}  \\
    732 &              4 &     0.036 &   -20.9 &    -9.85$\pm$0.07 &   -39.71$\pm$0.07 &    ....\hspace*{0.5cm} &    ....\hspace*{0.5cm}  \\
    301 &              2 &     0.036 &   -25.2 &   -47.86$\pm$0.05 &    24.89$\pm$0.11 &    ....\hspace*{0.5cm} &    ....\hspace*{0.5cm}  \\
    361 &              2 &     0.035 &    -5.3 &   117.76$\pm$0.07 &    -5.15$\pm$0.09 &    ....\hspace*{0.5cm} &    ....\hspace*{0.5cm}  \\
    300 &            2,3 &     0.035 &   -22.6 &   -49.35$\pm$0.04 &    20.23$\pm$0.06 &    ....\hspace*{0.5cm} &    ....\hspace*{0.5cm}  \\
    121 &              1 &     0.034 &   -27.1 &   -24.93$\pm$0.08 &   -29.66$\pm$0.07 &    ....\hspace*{0.5cm} &    ....\hspace*{0.5cm}  \\
     76 &              1 &     0.034 &   -28.8 &   -38.53$\pm$0.11 &    34.72$\pm$0.10 &    ....\hspace*{0.5cm} &    ....\hspace*{0.5cm}  \\
     40 &              1 &     0.034 &     3.4 &    22.05$\pm$0.08 &   -27.83$\pm$0.07 &    ....\hspace*{0.5cm} &    ....\hspace*{0.5cm}  \\
    344 &              2 &     0.032 &   -21.4 &   -37.76$\pm$0.05 &    -5.15$\pm$0.06 &    ....\hspace*{0.5cm} &    ....\hspace*{0.5cm}  \\
    674 &              4 &     0.032 &    -2.0 &    56.91$\pm$0.10 &     4.33$\pm$0.10 &    ....\hspace*{0.5cm} &    ....\hspace*{0.5cm}  \\
     74 &              1 &     0.032 &   -26.4 &    11.70$\pm$0.07 &   -34.92$\pm$0.06 &    ....\hspace*{0.5cm} &    ....\hspace*{0.5cm}  \\
     68 &              1 &     0.031 &     6.7 &   -74.72$\pm$0.09 &   -80.45$\pm$0.07 &    ....\hspace*{0.5cm} &    ....\hspace*{0.5cm}  \\
    355 &              2 &     0.028 &   -41.3 &   123.40$\pm$0.06 &   200.09$\pm$0.07 &    ....\hspace*{0.5cm} &    ....\hspace*{0.5cm}  \\
    133 &              1 &     0.027 &     6.0 &   -26.65$\pm$0.11 &   -53.43$\pm$0.09 &    ....\hspace*{0.5cm} &    ....\hspace*{0.5cm}  \\
     97 &              1 &     0.025 &    18.7 &   -42.42$\pm$0.17 &   -63.51$\pm$0.10 &    ....\hspace*{0.5cm} &    ....\hspace*{0.5cm}  \\
\end{longtable} 
\tablefoot{
\\
\tablefoottext{a}{The VLBA epochs are: 1)~March 6, 2023; \ 2)~June 5, 2023; \ 3)~August 5, 2023; \ 4)~October 19, 2023.} \\
Column~1 gives the feature label number; Col.~2 lists the observing epochs at which the feature was detected;
Cols.~3~and~4 provide the intensity of the strongest spot
and the intensity-weighted LSR velocity, respectively, averaged over the
observing epochs; Cols.~5~and~6 give the position offsets (with
the associated errors) along the RA and DEC axes, relative to feature~\#0, measured at the first epoch of detection; Cols.~7~and~8 give the components of the proper motion (with the associated errors) along the RA and DEC axes relative to feature~\#0.\\
The absolute position of the feature~\#0 at the first epoch on March 6, 2023, is: 
RA~(J2000) = 21$^{\rm h}$ 09$^{\rm m}$ 21\fs70580, DEC~(J2000) = 52\degree 22$^{\prime}$ 36\farcs9458, with an accuracy of \ $\pm$0.5~mas. 
}   

\end{document}